\documentclass[aps,twocolumn,superscriptaddress]{revtex4-1}

\usepackage{amsmath,amssymb,graphicx}
\usepackage{algorithmic}
\usepackage{enumerate}
\usepackage{color}
\graphicspath{{./fig-jpg/}{./fig-ps/}}

\begin{document}

\title{Hydrodynamics and two-dimensional dark lump solitons for polariton superfluids}

\author{D.~J. Frantzeskakis}
\affiliation{Department of Physics, National and Kapodistrian University of Athens, Panepistimiopolis, Zografos, Athens 15784, Greece}

\author{T.~P. Horikis}
\affiliation{Department of Mathematics, University of Ioannina, Ioannina 45110, Greece}

\author{A.~S.\ Rodrigues}
\affiliation{Departamento de F\'{\i}sica/CFP, Faculdade de Ci\^{e}ncias, Universidade do Porto, 
R. Campo Alegre, 687 - 4169-007 Porto, Portugal}

\author{P.~G. Kevrekidis}
\affiliation{Department of Mathematics and Statistics, University of Massachusetts, Amherst, Massachusetts 01003-4515 USA}

\author{R.~Carretero-Gonz{\'a}lez}
\affiliation{Nonlinear Dynamical Systems Group,
Computational Sciences Research Center, and
Department of Mathematics and Statistics,
San Diego State University, San Diego, California 92182-7720, USA}

\author{J. Cuevas-Maraver}
\affiliation{Grupo de F\'{i}sica No Lineal, Departamento de F\'{i}sica Aplicada I,
Universidad de Sevilla. Escuela Polit\'{e}cnica Superior, C/ Virgen de \'{A}frica, 7, 41011-Sevilla, Spain\\
Instituto de Matem\'{a}ticas de la Universidad de Sevilla (IMUS). 
Edificio Celestino Mutis. Avda. Reina Mercedes s/n, 41012-Sevilla, Spain}

\begin{abstract}

We study a two-dimensional incoherently pumped exciton-polariton condensate described by
an open-dissipative Gross-Pitaevskii equation 
for the polariton dynamics coupled to a rate equation for the exciton density.
Adopting a hydrodynamic approach, we use multiscale expansion methods to derive several
models appearing in the context of shallow water waves with viscosity. In particular,
we derive a Boussinesq/Benney-Luke type equation and its far-field expansion
in terms of Kadomtsev-Petviashvili-I (KP-I) equations for right-
and left-going waves. From the KP-I model, we predict the existence of
vorticity-free, weakly (algebraically) localized two-dimensional dark-lump solitons.
We find that, in the presence of dissipation, dark lumps exhibit a lifetime
three times larger than that of planar dark solitons. Direct numerical simulations
show that dark lumps do exist, and their dissipative dynamics is well captured by
our analytical approximation. It is also shown that
lump-like and vortex-like structures
can spontaneously be formed as a result of the transverse ``snaking'' instability of
dark soliton stripes.
\end{abstract}

%\begin{keyword}
%%
%Exciton-polariton condensates \sep dark solitons \sep lumps \sep
%open-dissipative Gross-Pitaevskii equation
%\sep Kadomtsev-Petviashvili equation \sep multiscale expansion methods


\maketitle

%\end{keyword}

%\end{frontmatter}

\section{Introduction}

Exciton-polariton superfluids, composed by hybrid
light-matter quasi-particles emerging in the regime of strong coupling,
offer  unique opportunities for studies on quantum, nonequilibrium
and nonlinear dynamics~\cite{rmp1st,cc}. Being intrinsically lossy ---and
hence being continuously replenished in order to be sustained---
polariton superfluids are
described by damped-driven versions of the Gross-Pitaevskii (GP) equation
\cite{wc,wc2,wc3,berl1,berl2} (see also the review of Ref.~\cite{cc}).
Such models have been successfully used for the theoretical study of
fundamental nonlinear phenomena that have been observed in experiments, e.g.,
the formation and dynamics of quantized vortices~\cite{vort1,vort2,vort3,vort3a}
and dark solitons~\cite{ds0,ds1,ds2,ds3,ds4,ds5,ds6}.

There are numerous works that have studied the dynamics of dark solitons in polariton 
condensates.
%~\cite{dsth1,dsth1b,dsth1c,dsthwe1,dsthwe2,ofy,dsthv1,dsthv2,wepla}.
For instance, dark solitons in polariton condensates coherently and 
resonantly driven by a pumping laser were studied in Refs.~\cite{dsth1,dsth1b,dsth1c}. 
Also, in the presence of
nonresonant pumping, simplified Ginzburg-Landau models~\cite{berl1,berl2} were
used to describe one-dimensional (1D) dark solitons and two-dimensional (2D) 
ring dark solitons in Refs.~\cite{dsthwe1,dsthwe2}.
In the same case (of nonresonant pumping),
and using the open-dissipative GP model of Refs.~\cite{wc,wc2,wc3},
which involves the coupling of polaritons to the exciton reservoir,
dark polariton solitons were analyzed using an
adiabatic approximation~\cite{ofy} and variational techniques~\cite{dsthv1,dsthv2}.
Finally, in Ref.~\cite{wepla}, the 1D open-dissipative GP model
was asymptotically reduced to an effective Korteweg-de Vries (KdV) equation with linear loss,
which was then used to describe dark soliton dynamics in polariton superfluids.

The recent work of Ref.~\cite{wepla} suggests a number of interesting questions.
Before asking some of these, it is relevant to mention the following.
The KdV equation is known to be a universal model describing shallow
water waves \cite{ablowitz2}, as well as ion-acoustic solitons in plasmas \cite{Infeld},
solitons in mechanical and electrical lattices, and so on \cite{Rem}. The KdV equation
describes uni-directional propagation and stems, as the far-field limit, from
bi-directional models appearing in various contexts ---predominantly in shallow water waves---
such as the Boussinesq \cite{ablowitz2,Infeld,Rem} and the Benney-Luke (BL) \cite{BL} equations.
Additionally, the generalization of the KdV in the 2D setting, namely
the Kadomtsev-Petviashvili (KP) equation, also stems from the 2D versions of
Boussinesq- and BL-type models~\cite{ablowitz2}. Then, one can ask whether
these models may be relevant to the context of polariton superfluids as well.
If yes, one can hope to use them towards predicting dynamical features of
the solitons, towards identifying
novel ---e.g., 2D--- solitonic structures in exciton-polariton systems
and, finally, towards quantifying the role of the open-dissipative nature
of these systems.
%And if yes,
%what is the role of the open-dissipative nature of these systems on the 2D soliton dynamics?

The scope of this work is to address these questions. In particular, our starting
point is the open-dissipative GP equation of Refs.~\cite{wc,wc2,wc3} in 2D, which is
perhaps the most customary approach for describing an incoherently pumped exciton-polariton
BEC. This model is composed by a dissipative GP equation for the macroscopic
wavefunction of the polariton condensate, coupled with a rate equation
for the exciton reservoir density. Adopting a hydrodynamic description,
we use multiscale expansion methods to derive ---under certain physically relevant conditions---
asymptotic reductions of the open-dissipative GP equation. Specifically,
at an intermediate stage of the asymptotic analysis, we obtain a Boussinesq/BL-type
equation with linear loss ---similarly to the case of shallow water waves when
viscosity is taken into account \cite{dut,jiang}. Next, we consider the far-field of
the Boussinesq/BL-type equation, and derive a pair of Kadomtsev-Petviashvili (KP)
equations \cite{KP} with linear loss, for right- and left-going waves. Such
KP models
(and also their 1D, KdV, counterpart) perturbed by a linear loss term have
been used
in shallow water wave settings involving straits of nonuniform water depth
\cite{newell,david}. It is important to point out that,
given the self-defocusing nature of the nonlinearity,
the derived KP model
is of the KP-I type, i.e., it is characterized by positive dispersion,
and arises in the context of shallow water waves, or in liquid thin films,
when surface tension dominates gravity \cite{ablowitz2}.

Next, in the absence of dissipation, 1D line soliton and 2D lump solutions
of the KP-I equation \cite{MJA} are used for the construction of two different types of
(approximate) soliton solutions of the original open-dissipative GP model:
(i) planar dark solitons, satisfying the 1D (KdV) counterpart of the KP-I ---similar
to the ones studied in the 1D setting of Ref.~\cite{wepla}--- and (ii) weakly localized
(i.e., algebraically decaying), vorticity-free, 2D dark solitons. The fact that
the underlying KP equation is of the KP-I type has important consequences:
since line solitons (lumps) of the KP-I are unstable (stable)
\cite{zakhpr} then, in the original GP model, planar dark solitons are
also unstable in 2D ---as was also demonstrated in the simulations of Ref.~\cite{ofy}--- 
while 2D dark lump solitons are dynamically robust. Importantly, similar 
dark lump solitons have been predicted and studied in nonlinear optics \cite{d1,d2,kod1}, 
atomic Bose-Einstein condensates (BECs) \cite{huang} (see also the reviews \cite{jpa}),
superfluid Fermi gases \cite{fermi}, and in laser-plasma interactions \cite{plasma}.
As was shown, these structures can emerge as a result of the slowly developing
transverse (alias ``snaking'') instability of the 1D dark solitons \cite{smirnov1}, or
during vortex-antivortex annihilation \cite{verma} (see also relevant work in
Refs.~\cite{smirnov2,smirnov3,helm}).

We also study the role of dissipation on the soliton dynamics, which is
particularly relevant due to the open-dissipative nature of the present
system. It is found that the amplitude of both the line solitons and the lumps 
decays exponentially in time, with a rate which is set by physical parameters of 
the problem, namely the polariton decay rate and the relative deviation of the 
uniform pumping from its threshold value. Remarkably, it is found that the lifetime 
of the weakly-localized dark lump soliton is three times larger than the one 
of the line soliton. This suggests that these structures have a good chance 
to be observed in experiments. Finally, we use direct numerical simulations 
to show that dark lumps do exist, and their dissipative dynamics is well described by
our asymptotic approach. In addition, the use of direct simulations shows that 
relevant lump ---as well as vortical--- structures can spontaneously be formed as 
a result of the transverse ``snaking'' instability of weak dark soliton stripes.
% PGK: As I mentioned I am especially hesitant to make this kind
% of statements --- it would certainly be useful to have phase info...

The paper is structured as follows. In Section~2, we present the model and use asymptotic
expansion methods to derive the effective Boussinesq/BL and KP-I equations. In Section~3,
employing the soliton solutions of the KP-I model, we construct corresponding approximate
soliton solutions of the open-dissipative GP equation. We also present results of
direct numerical simulations depicting: (i) the dissipative dynamics of the dark lumps,
as well as (ii) the snaking instability of dark soliton stripes.
Finally, Section~4 summarizes our conclusions and provides directions for relevant future work.

\section{The model and its analytical consideration}
\label{sec:model}

\subsection{The open-dissipative Gross-Pitaevskii model}

Let us consider a 2D incoherently pumped (far from resonance) exciton-polariton condensate
described, in the mean-field approximation, by a generalized open-dissipative GP system
where the polariton wavefunction $\Psi(\boldsymbol{r},t)$ is coupled to a rate equation 
for the exciton reservoir density $n(\boldsymbol{r},t)$~\cite{wc,wc2,wc3}:
\begin{eqnarray}
i\hbar \Psi_t
%\frac{\partial \Psi}{\partial t}
&=& \left[ -\frac{\hbar^2}{2M}\Delta
%\Psi
+g_C |\Psi|^2
%\Psi
+ g_R n
%\Psi
+ \frac{i\hbar}{2}(Rn-\gamma_C)\right]\Psi,~~~~
\label{Psi} \\
%
%\frac{\partial n_R}{\partial t}
n_t &=& P(\boldsymbol{r},t)-(\gamma_R+R|\Psi|^2)n,
\label{nR}
\end{eqnarray}
where subscripts in the fields $\Psi$ and $n$ denote partial derivatives
and $\Delta \equiv \partial_x^2+\partial_y^2$ is the 2D Laplacian. 
In these equations, the polaritons, with effective mass (lower polariton branch)
$M$, have a (self) nonlinear interaction strength $g_C$ and are coupled,
with coupling strength $g_R$, to the exciton reservoir.
Furthermore, $R$ measures the reservoir's rate of stimulated scattering, while
$\gamma_C$ and $\gamma_R$ are the polariton and exciton loss rates, respectively.
Finally, $P(\boldsymbol{r},t)$ is the exciton creation rate induced
by the spatio-temporal laser pumping profile.
It is important to note that within this model, the polariton condensate includes
an intrinsic repulsive (defocusing) nonlinearity ($g_C>0$).
%, stemming from the repulsive interaction between excitons~\cite{rmp1st,cc}.

To simplify our analysis, we express Eqs.~(\ref{Psi})-(\ref{nR}) in dimensionless form by:
scaling space in terms of the healing length $\bar\xi=\hbar/\sqrt{Mg_C n_C}$
(where $n_C$ is the background's condensate density), scaling time in units of
$t_0=\bar\xi/c_S$ (where $c_S=\sqrt{g_C n_C/M}$ is the speed of sound),
and scaling densities (namely $|\Psi|^2$ and $n$) in terms of $n_C$.
Using these scalings yields
\begin{eqnarray}
%i\frac{\partial \Psi}{\partial t}
i\Psi_t&=& -\frac{1}{2}\Delta \Psi
+|\Psi|^2\Psi + g_R n \Psi + \frac{i}{2}(Rn-\gamma_C)\Psi,
\label{Psin} \\
%\frac{\partial n_R}{\partial t}
n_t&=& P(\boldsymbol{r},t)-(\gamma_R+R|\Psi|^2)n,
\label{nRn}
\end{eqnarray}
where now $g_R$ and $R$ are measured in units of $g_C$ and $g_C/\hbar$, respectively,
$\gamma_C$ and $\gamma_R$ are measured in units of $1/t_0$, and the laser pump $P(x,t)$
is measured in units of $n_C/t_0$.

The starting step to describe the polariton condensate as a fluid is to employ
the so-called Madelung transformation $\Psi = \sqrt{\rho}\,\exp(i\varphi)$
that expresses the polariton evolution in terms of its density $\rho$
and phase $\varphi$. 
Then, after separating real and imaginary parts, one obtains the fluid-like equations:
\begin{eqnarray}
&&
%\frac{\partial \varphi}{\partial t}
\varphi_t +\rho+\frac{1}{2}\left(\boldsymbol{\nabla} \phi\right)^2
-\frac{1}{2}\rho^{-1/2}\Delta \rho^{1/2}+g_R n=0,
\label{phit} \\
&&
%\frac{\partial \rho}{\partial t}
\rho_t+\boldsymbol{\nabla} \cdot (\rho \boldsymbol{\nabla}\phi)
-(R n -\gamma_C)\rho=0,
\label{rhot} \\
&&
%\frac{\partial n}{\partial t}
n_t-P+(\gamma_R +R \rho)n =0,
\label{nRt}
\end{eqnarray}
where $\boldsymbol{\nabla} \equiv (\partial_x,~\partial_y)$ is the gradient operator.
From this point onward, we restrict our attention to the constant-in-time and spatially uniform
pumping profile $P(x,t)=P_0$. For this pumping profile the spatially homogeneous 
steady-states of the exciton-polariton system correspond to
\begin{eqnarray}
\rho=\rho_0, \quad n = n_0, \quad \varphi=-\mu t,
\label{ss}
\end{eqnarray}
where the steady-state condensate and reservoir background densities, $\rho_0$ and $n_0$, 
as well as the chemical potential $\mu$, are given by \cite{wc,ofy,wepla}:
\begin{eqnarray}
\rho_0=\frac{P_0-P_0^{({\rm th})}}{\gamma_C},
~~
n_0=\frac{\gamma_C}{R},
~~
\mu = \rho_0 + \frac{g_R \gamma_C}{R},
\label{n0}
\end{eqnarray}
where we have defined
\begin{eqnarray}
P_0^{({\rm th})}\equiv \frac{\gamma_R\gamma_C}{R}.
\end{eqnarray}
In the above, the inequality $P_0>P_0^{({\rm th})}$ must be satisfied for the
polariton density to be meaningful (i.e., positive). This implies that a non-zero
polariton steady-state can only be sustained provided that the pump strength
$P_0$ exceeds the threshold value $P_0^{({\rm th})}$ given above. Therefore,
by defining the following positive parameter, corresponding to the relative 
deviation of the pumping from the threshold, 
\begin{equation}
\alpha=\frac{P_0-P_0^{({\rm th})}}{P_0^{({\rm th})}},
\label{alpha}
\end{equation}
we can express the equilibrium condensate density as $\rho_0=(\gamma_R/R)\alpha$.

In what follows, we consider the case $\gamma_C \ll \gamma_R$ corresponding to the
physically realistic scenario whereby the exciton reservoir follows adiabatically
the polariton condensate evolution \cite{wc}.
In order to employ a multiscale expansion approach, we follow the smallness of the
relevant quantities by introducing the formal small parameter $0<\epsilon \ll 1$
and assume that $\gamma_C=\epsilon \tilde{\gamma}_C$, where $\tilde{\gamma}_C$
and $\gamma_R$ remain of order $\mathcal{O}(1)$.
By the same token, we consider that the reservoir's scattering rate $R$ and the 
relative deviation of the pumping from the threshold, $\alpha$, are also
relatively small \cite{ofy} and of order $\epsilon$: $R = \epsilon \tilde{R}$ and
$\alpha = \epsilon \tilde{\alpha}$, where $\tilde{R}$ and $\tilde{\alpha}$ are
of order $\mathcal{O}(1)$. All other parameters and relevant quantities
---such as the densities $\rho_0$ and $n_0$, the pump threshold $P_0^{({\rm th})}$,
and the chemical potential $\mu$--- are assumed to be of order $\mathcal{O}(1)$.

\subsection{Effective nonlinear evolution equations }

\subsubsection{The Boussinesq/Benney-Luke--type equation}

To better underline the hydrodynamic origin of the soliton solutions 
presented below, we will first derive a Boussinesq/Benney-Luke--type equation.
We thus seek solutions of Eqs.~(\ref{phit})-(\ref{nRt}) in the form of the
following asymptotic expansions:
\begin{subequations}
\begin{eqnarray}
\phi&=&-\mu t +\epsilon^{1/2}\Phi,
\label{expansions1} \\
%\rho=1+\sum_{j=1}^{\infty}\vareps^j \rho_j,
\rho&=&\rho_0+\epsilon \rho_1 + \epsilon^2 \rho_2 + \cdots,
\label{expansions2} \\
n&=& n_0+ \epsilon^2 n_1 + \epsilon^3 n_2  + \cdots,
%\sum_{j=1}^{\infty}\vareps^j n_j,
\label{expansions3}
\end{eqnarray}
\end{subequations}
where $\epsilon$ is the same formal small parameter %($0<\epsilon \ll 1$)
used for the scaling of the parametric dependences above, while the
unknown real functions $\Phi$, $\rho_j$ and $n_j$ ($j=1,2,\ldots$) are assumed to
depend on the slow variables:
\begin{equation}
X=\epsilon^{1/2}x, \quad Y= \epsilon^{1/2}y, \quad T=\epsilon^{1/2}t.
\label{slow}
\end{equation}
Substituting the expansions (\ref{expansions1})-(\ref{expansions3})
into Eqs.~(\ref{phit})-(\ref{nRt}), and
using the variables in Eq.~(\ref{slow}), we obtain the following results.
First, up to order $\mathcal{O}(\epsilon)$, Eq.~(\ref{phit}) reads:
\begin{eqnarray}
\Phi_T+\rho_1+\epsilon\left[\frac{1}{2}(\tilde{\boldsymbol{\nabla}}\Phi)^2
-\frac{1}{4}\tilde{\Delta}\rho_1 +\rho_2 \right]=0,
%= \mathcal{O}(\epsilon^2),
\label{Phi}
\end{eqnarray}
where $\tilde{\Delta} = \partial_X^2 + \partial_Y^2$ and
$\tilde{\boldsymbol{\nabla}}=(\partial_X,~\partial_Y)$.
Second, Eq.~(\ref{rhot}) leads, at orders $\mathcal{O}(\epsilon^{3/2})$ and
$\mathcal{O}(\epsilon^{5/2})$, to the following equations, respectively:
\begin{eqnarray}
&&\rho_{1T}+\rho_0\tilde{\Delta}\Phi=0,
\label{nn1}
\\
&&\rho_{2T}+\tilde{\boldsymbol{\nabla}} \cdot (\rho_1 \tilde{\boldsymbol{\nabla}}\Phi)=0.
\label{nn2}
\end{eqnarray}
Third, Eq.~(\ref{nRt}), at the leading order in $\epsilon$, i.e., at order
$\mathcal{O}(\epsilon^{2})$, yields:
\begin{equation}
n_1 = -\frac{\tilde{\gamma}_C}{\gamma_R} \rho_1,
\label{nqconn1}
\end{equation}
connecting the reservoir density $n_1$ to the polariton density $\rho_1$. Obviously,
once $\rho_1$ is found, then $n_1$ and $\Phi$ can be respectively derived from
Eq.~(\ref{nqconn1}) and the leading-order part of Eq.~(\ref{Phi}), namely $\Phi_T+\rho_1=0$.

At the present order of approximation, Eqs.~(\ref{Phi})-(\ref{nn2})
do not incorporate dissipative terms. The lowest-order such term appears in Eq.~(\ref{nn2}),
and has the form $-\epsilon^3 \tilde{\alpha}\gamma_R n_1$, i.e., it is a
term of order $\mathcal{O}(\epsilon^3)$. To take into account this term,
we may modify Eq.~(\ref{nn2}) by adding to its right-hand side the additional
term $-\epsilon^{1/2} \tilde{\alpha}\gamma_R n_1$.
Taking into regard this modification,
we may proceed as follows. Using Eq.~(\ref{nn1}) and the modified Eq.~(\ref{nn2}),
we can eliminate the functions $\rho_{1,2}$ from Eq.~(\ref{Phi}) and derive the following
equation
for $\Phi$:
\begin{eqnarray}
\!\!\!\!\!\!\!
&&\Phi_{TT}-C^2 \tilde{\Delta}\Phi
+\epsilon\left[\frac{1}{4}\tilde{\Delta}^2 \Phi
+ \frac{1}{2}\partial_T (\tilde{\boldsymbol{\nabla}}\Phi)^2
+ \tilde{\boldsymbol{\nabla}} \cdot (\Phi_T \tilde{\boldsymbol{\nabla}}\Phi)
\right] \nonumber \\
%\right. \nonumber \\
%&&\left.
%+ \frac{1}{2}\partial_T (\tilde{\boldsymbol{\nabla}}\Phi)^2
%+ \tilde{\boldsymbol{\nabla}} \cdot (\Phi_T \tilde{\boldsymbol{\nabla}}\Phi)\right]
\!\!\!\!\!\!\!
&&+\epsilon^{3/2} \tilde{\alpha}\tilde{\gamma}_C \Phi_T =0,
\label{Bou}
\end{eqnarray}
where the squared velocity $C^2$ is given by:
\begin{equation}
C^2 =\rho_0,
\end{equation}
up to corrections of $\mathcal{O}(\epsilon^2)$.

It is clear that, to leading-order, Eq.~(\ref{Bou})
is a linear wave equation. In addition, at the order $\mathcal{O}(\epsilon)$, Eq.~(\ref{Bou})
incorporates fourth-order dispersion terms and quadratic nonlinear terms.
Obviously, Eq.~(\ref{Bou}) resembles the Boussinesq and Benney-Luke \cite{BL} equations,
which describe bidirectional shallow water waves, in the framework of small-amplitude
and long-wavelength approximations \cite{ablowitz2}; note that such Boussinesq-type
models have also been used in other contexts, ranging from ion-acoustic waves
in plasmas \cite{Infeld} to mechanical lattices and electrical transmission lines \cite{Rem}.
Finally, at the order $\mathcal{O}(\epsilon^{3/2})$, Eq.~(\ref{Bou}) also includes
a dissipative term (see below) proportional to $\tilde{\alpha}\tilde{\gamma}_C$, i.e.,
depending on the polariton decay rate and the relative deviation of the uniform
pumping from its threshold value. Such a Boussinesq-like model with constant dissipation
can also be derived in the context of shallow water waves, upon incorporating a
dissipative term ---to account for the presence of viscosity--- in the free-surface dynamical
boundary condition \cite{dut} (see also Ref.~\cite{jiang}).

Before proceeding further, it is relevant to focus, at first, on the linear dispersion
relation of Eq.~(\ref{Bou}), which can be derived as follows. Seeking small-amplitude
solutions of Eq.~(\ref{Bou}) behaving like
$\Phi \propto \exp[i(\boldsymbol{k}\cdot \boldsymbol{r} -\omega t)]$,
with $\boldsymbol{r}=(x,~y)$, we find that the perturbations'
wavevector $\boldsymbol{k}=(k_x,~k_y)$ and frequency $\omega$ obey the dispersion
relation:
\begin{equation}
\omega(|\boldsymbol{k}|)=\pm \sqrt{\omega_B^2(|\boldsymbol{k}|) + \left( \frac{1}{2}\epsilon^{3/2}
\tilde{\alpha}\tilde{\gamma}_C\right)^2} -\frac{i}{2}\epsilon^{3/2} \tilde{\alpha}\tilde{\gamma}_C,
\label{drb}
\end{equation}
% PGK: are you sure about the ``+'' inside the radical ??
% I would think that because of the ``i'' it is a (-) rather than (+)
%
where $\omega_B^2(|\boldsymbol{k}|) = |\boldsymbol{k}|^2 C^2
+(1/4) \epsilon |\boldsymbol{k}|^4$ %(with $|\boldsymbol{k}|^2 =k_x^2+k_y^2$)
is the standard Bogoliubov dispersion relation for a condensate at equilibrium \cite{jpa}.
It is clear that Eq.~(\ref{drb}) suggests a decay rate of linear waves
proportional to $\tilde{\alpha}\tilde{\gamma}_C$. Below we will show that
localized nonlinear waves, in the form of 1D line solitons and 2D lumps,
satisfying a KP-I equation that will be derived as the far field of Eq.~(\ref{Bou}),
also feature a decay rate proportional to $\tilde{\alpha}\tilde{\gamma}_C$.
This can also be suggested by the linear theory as follows.
Using $|\boldsymbol{k}|^2 =k_x^2+k_y^2$, and keeping terms up to the order
$\mathcal{O}(\epsilon^{2})$, we cast Eq.~(\ref{drb})
into the form:
\begin{eqnarray}
\omega \approx &\pm& C k_x
\left(1+\frac{k_y^2}{k_x^2}\right)^{1/2}
\left[1+\frac{\epsilon}{4C^2}
k_x^2 \left(1+\frac{k_y^2}{k_x^2}\right)\right]^{1/2} \nonumber \\
&-&\frac{i}{2}\epsilon^{3/2} \tilde{\alpha}\tilde{\gamma}_C,
\label{lwl1}
\end{eqnarray}
with $\pm$ corresponding to right- and left-going waves.
Then, considering a quasi-2D evolution with $k_y/k_x =\mathcal{O}(\epsilon^{1/2})$,
we may further simplify Eq.~(\ref{lwl1}) as follows:
\begin{equation}
\frac{1}{C}\omega k_x = \pm \left[ k_x^2 + \frac{\epsilon}{8C^2} k_x^4 +\frac{\epsilon}{2}k_y^2\right]
-\frac{i}{2C}\epsilon^{3/2} \tilde{\alpha}\tilde{\gamma}_C k_x.
\label{lwl2}
\end{equation}
Then, using $\omega \rightarrow i\partial_t$, $k_{x,y} \rightarrow -i\partial_{x,y}$,
it is found that the linear PDE associated with this dispersion relation is:
$\partial_x [\pm q_t+ C q_x - (\epsilon/8C)q_{xxx}] +(\epsilon C/2)q_{yy}=
\mp \epsilon^{3/2} \tilde{\alpha}\tilde{\gamma}_C q_x$. To this end,
employing the transformation $x'=x-Ct$ and using the slow time $t'=\epsilon t$, the above equation
takes the form
\begin{equation}
\partial_{x'}\left(\pm q_{t'}-\frac{1}{8C} q_{x'x'x'}\right)+\frac{C}{2}q_{yy}=
\mp \frac{1}{2}\epsilon^{1/2} \tilde{\alpha}\tilde{\gamma}_C q_{x'},
\label{lKP}
\end{equation}
which is a linear KP equation incorporating, in its right-hand side,
a small [of order $\mathcal{O}(\epsilon^{1/2})$] linear loss term.
To derive the full nonlinear version of the KP model, in the next section, we resort to
the method of multiple scales.

\subsubsection{The Kadomtsev-Petviashvili-I equation}

We now proceed to derive the far-field equations stemming from Eq.~(\ref{Bou}),
in the framework of multiscale asymptotic expansions. As is well known,
the far-field of the Boussinesq equation in $(1+1)$-dimensions
is a pair of two KdV equations \cite{ablowitz2}, while in $(2+1)$-dimensions,
it is a pair of KP equations \cite{peli1,hor1,hor2}, for right- and left-going waves.
The KP equation, can be derived under the additional
assumptions of quasi-two-dimensionality and unidirectional
propagation. In particular, first we introduce the asymptotic
expansion:
\begin{equation}
\Phi = \Phi_0 + \epsilon \Phi_1 + \cdots,
\label{expphi}
\end{equation}
where the unknown functions $\Phi_{\ell}$ ($\ell=0,1,\ldots$) depend on the variables
\begin{equation}
\xi=X-CT,~~\eta=X+CT,~~\mathcal{Y}=\epsilon^{1/2}Y,~~\mathcal{T}=\epsilon T.
\notag
\end{equation}
Substituting this expansion into Eq.~(\ref{Bou}), at the leading-order in
$\epsilon$, we obtain the wave equation
\begin{equation}
\Phi_{0\xi \eta}=0,
\end{equation}
implying that $\Phi_0$ can be expressed as a superposition of a right-going wave,
$\Phi_0^{(R)}$, depending on $\xi$, and a left-going one, $\Phi_0^{(L)}$, depending on $\eta$,
namely:
\begin{equation}
\Phi_{0}=\Phi_0^{(R)}(\xi,\mathcal{Y},\mathcal{T})+\Phi_0^{(L)}(\eta,\mathcal{Y},\mathcal{T}).
\label{rl}
\end{equation}
In addition, at order $\mathcal{O}(\epsilon)$, and taking into regard ---as before---
the correction including the dissipative term of order $\mathcal{O}(\epsilon^{3/2})$
[cf.~last term in the left-hand side of Eq.~(\ref{Bou})], we obtain the equation:
\begin{widetext}
\begin{eqnarray}
4C^2\Phi_{1\xi\eta} = &-&C\left(\Phi_{0\xi\xi}^{(R)}\Phi_{0\eta}^{(L)}
-\Phi_{0\xi}^{(R)}\Phi_{0\eta\eta}^{(L)} \right) \nonumber \\
&+&\left[\partial_{\xi}
\left(-2C\Phi_{0\mathcal{T}}^{(R)} +\frac{1}{4}\Phi_{0\xi\xi\xi}^{(R)}
-\frac{3C}{2}\Phi_{0\xi}^{(R)2}
-\epsilon^{1/2} C \tilde{\alpha}\tilde{\gamma}_C \Phi_0^{(R)} \right)
-C^2\Phi_{0\mathcal{Y}\mathcal{Y}}^{(R)}\right]
\nonumber \\
&+& \left[\partial_{\eta}\left(2C\Phi_{0\mathcal{T}}^{(L)}
+\frac{1}{4}\Phi_{0\tilde{\eta}\tilde{\eta}\tilde{\eta}}^{(L)}
+\frac{3C}{2}\Phi_{0\tilde{\eta}}^{(L)2}
+\epsilon^{1/2} C \tilde{\alpha}\tilde{\gamma}_C \Phi_0^{(L)}
\right)-C^2\Phi_{0\mathcal{Y}\mathcal{Y}}^{(L)}
\right].
%\nonumber
\label{phif}
\end{eqnarray}
\end{widetext}
When integrating Eq.~(\ref{phif}), secular terms arise from the square brackets
in the right-hand side, which are functions of $\xi$ or $\eta$ alone, not both.
Removal of these secular terms leads to two uncoupled nonlinear evolution equations for
$\Phi_0^{(R)}$ and $\Phi_0^{(L)}$. Furthermore, using the equation
$\Phi_T=-\rho_1$, obtained from the leading-order part of Eq.~(\ref{Phi}), the
amplitude $\rho_1$ is also decomposed to a left- and a right-going wave, i.e., $\rho_1 =
\rho_1^{(R)}+\rho_1^{(L)}$, with:
\begin{equation}
C \Phi_{0 \xi}^{(R)}=\rho_1^{(R)}, \quad C
\Phi_{0 \eta}^{(L)}=-\rho_1^{(L)}.
\label{phixi}
\end{equation}
To this end, using the above equations for $\Phi_0^{(R)}$ and $\Phi_0^{(L)}$, along with
Eqs.~(\ref{phixi}), yields the following KP equations for $\rho_1^{(R,L)}$:
\begin{eqnarray}
&&\partial_{\mathcal{X}}\left(\pm \rho_{1\mathcal{T}}^{(R,L)}
-\frac{\alpha}{8C}\rho_{1\mathcal{X}\mathcal{X}\mathcal{X}}^{(R,L)}
+\frac{3C}{2}\rho_{1}^{(R,L)} \rho_{1\mathcal{X}}^{(R,L)}\right)
\nonumber \\
&&+\frac{1}{2}C\rho_{1\mathcal{Y}\mathcal{Y}}^{(R,L)}=
\mp \frac{1}{2}\epsilon^{1/2} \tilde{\alpha}\tilde{\gamma}_C \rho_{1\mathcal{X}}^{(R,L)},
%\nonumber \\
\label{KP-I}
\end{eqnarray}
where $\mathcal{X}=\xi$ or $\mathcal{X}=\eta$, as well as $\pm$, corresponds
to the right (R)- or the left (L)-going wave.
Obviously, the above equations are of the KP type, and incorporate a dissipative perturbation
having the form of a linear loss term. Generally, the KP equation is a 2D extension
of the KdV equation ---cf.~Eq.~(\ref{KP-I}) for $\partial_\mathcal{Y}=0$. The particular
form of Eq.~(\ref{KP-I}) is of the KP-I type, i.e., it is characterized by positive dispersion,
and is known to govern shallow water waves, in the case where surface tension dominates
gravity \cite{ablowitz2}.

Importantly, the KP-I equation is known to display the effect of transverse instability
and self-focusing of planar (quasi-1D)
localized structures, so-called line solitons
(cf.~next section). In particular, as was first shown in hydrodynamics and plasma
physics \cite{zakhpr}, line solitons develop undulations and
eventually decay into lumps \cite{infeld}. Additionally, in optics, the
asymptotic reduction of the defocusing 2D nonlinear Schr\"odinger (NLS)
equation to KP-I \cite{kuz1,peli1}, and the
instability of the line solitons of the latter, was used
to better understand the transverse instability of rectilinear
dark solitons: indeed, these structures being subject to transverse (alias ``snaking'')
instability, also develop undulations and eventually decay into vortex pairs \cite{peli1,pelirev}
or, in some cases, into 2D vorticity-free structures resembling KP lumps \cite{smirnov1}.
A recent analysis of the resulting line soliton filament dynamics can be found in Ref.~\cite{AI}.

Below we will present both the unstable 1D solitons and the stable 2D solitons of
the KP-I model, namely the lumps. We will focus on the latter, and show that the
lump solution of the KP-I equation can be used to construct
weakly-localized 2D dark solitons of the original model.

\section{Soliton solutions}

\subsection{Unperturbed soliton solutions}

% PGK: Dimitri, I am now quite confused... On the
% one hand, it is said that we examine the unperturbed dynamics
% for \epsilon=0 and then you proceed to use $\epsilon$ all over
% the place, in (31), (32), etc. This does not seem consistent...
% If you want to set the dissipation to 0 it's a different story
% but it's not obvious that you can set \epsilon=0...

Without loss of generality, let us consider the case of right-going waves and
study, at first, the unperturbed version of the KP-I equation~(\ref{KP-I}),
%for $\epsilon=0$.
corresponding to the case where dissipation is absent.
We can use the soliton solutions of this reduced
model, which we present below, to find
approximate soliton solutions of the open dissipative GP Eqs.~(\ref{Psin})-(\ref{nRn}).
Indeed, in terms of the original (dimensionless) variables and coordinates,
$x$, $y$ and $t$, one may write down an approximate [up to order
$\mathcal{O}(\epsilon)$] solution for the macroscopic wavefunction
$\Psi$ of the polariton condensate and the exciton density $n_R$ as follows:
\begin{eqnarray}
\Psi &\approx& \sqrt{\rho_0+\epsilon \rho_{1}^{(R)}}
\exp\left(-i\mu t + i\epsilon^{1/2} \Phi_{0}^{(R)}\right),
\label{psis} \\
n &\approx& n_0 - \epsilon \frac{\gamma_C}{\gamma_R} \rho_{1}^{(R)},
\label{ns}
\end{eqnarray}
where $\rho_{1}^{(R)}$ is a soliton of KP-I~(\ref{KP-I}) and $\Phi_{0}^{(R)}$ is
the respective phase, which can be directly  found from the first of Eqs.~(\ref{phixi}).

Let us now present the soliton solutions of the KP-I, Eq.~(\ref{KP-I}), which can
be distinguished into two types. The first one which is
quasi-1D, and is usually called ``line soliton'' \cite{ablowitz2}, has the form:
\begin{equation}
\rho_{1}^{(R)}=
%(\xi,\mathcal{Y},\mathcal{T})=
-\kappa^2{\rm sech}^2 Z, \quad Z=\kappa[\mathcal{X}-\zeta(\mathcal{T})],
\label{1ds}
\end{equation}
where $\kappa$ is a free parameter linking the soliton's amplitude to its velocity,
$\zeta(\mathcal{T})=4\kappa^2 \mathcal{T} + \zeta_0$ is the soliton center
(with the constant $\zeta_0$ denoting the initial soliton location), and
$d\zeta/d\mathcal{T}=4\kappa^2$ is the soliton velocity in the
$(\xi,\mathcal{T})$ reference frame. The above 1D structure is actually
the soliton solution of the KdV equation associated with the KP-I Eq.~(\ref{KP-I})
extended uniformly in the $\mathcal{Y}$ direction.
The phase $\Phi_0^{(R)}$ associated to this solution can be
obtained from the first of Eqs.~(\ref{phixi}):
\begin{equation}
\Phi_0^{(R)} = -\frac{\kappa^2}{C}\tanh Z,
\label{1dsph}
\end{equation}
and it should be mentioned that, in terms of the original (dimensionless) coordinates,
$x$, $y$ and $t$, the variable $Z$ reads:
\begin{equation}
Z=\epsilon^{1/2} \kappa \left[x-\left(C-\frac{\epsilon\kappa^2}{2C} \right)t-x_0 \right].
\end{equation}
Clearly, in this case, the solution~(\ref{psis}) has the form of a sech-shaped
density dip, with a tanh-shaped phase jump across the density minimum, and it
is thus a dark (gray) soliton. On the other hand, the exciton
density~(\ref{ns}) follows the form of an antidark soliton soliton, i.e., it has a
sech$^2$ hump shape on top of the background, at the location of the dark polariton soliton,
and asymptotes (for $x\rightarrow \pm \infty$) to the equilibrium density $n_0$.
Notice that the dynamics of this solution was studied systematically
in the context of the 1D analogue of the open-dissipative GP model~(\ref{Psin})-(\ref{nRn})
in the recent work of Ref.~\cite{wepla}. Furthermore, in the 2D setting, the transverse instability of
1D dark solitons of a similar form was also studied in the context of exciton-polariton
condensates in Ref.~\cite{ofy}.

Let us next proceed with the second type of soliton solution of Eq.~(\ref{KP-I}),
which is of primary interest herein. This is a genuinely 2D soliton,
known as ``lump'' \cite{ablowitz2}, and is of the form:
\begin{equation}
\rho_{1}^{(R)}
%(\xi,\mathcal{Y},\mathcal{T})
=-2
\frac{-\left(\xi+\frac{3V}{2C} \mathcal{T}\right)^2+\frac{3W}{C^2}\mathcal{Y}^2+\frac{1}{4\beta^2}}{\left[\left(\xi+\frac{3V}{2C} \mathcal{T}\right)^2+\frac{3W}{C^2}\mathcal{Y}^2+\frac{1}{4\beta^2}\right]^2},
\label{lumpo}
\end{equation}
where $\beta$ is a free parameter connecting the velocity ($-\frac{3V}{2C}$) 
and the inverse width through $V=W=\beta^2$ and thus linking the soliton amplitude with
its velocity and transverse width. This solution
is weakly localized, since it decays algebraically as
$(\mathcal{\xi}^2+\mathcal{Y}^2)^{1/2} \rightarrow \infty$.
Employing, as before, the first of Eqs.~(\ref{phixi}), we can also obtain
the associated phase $\Phi_0^{(R)}(\xi,\mathcal{Y},\mathcal{T})$:
\begin{eqnarray}
\Phi_0^{(R)}
%(\xi,\mathcal{Y},\mathcal{T})
= -\frac{2}{C} \left[\frac{\xi+\frac{3V}{2C} \mathcal{T}}
{\left(\xi+\frac{3V}{2C} \mathcal{T}\right)^2+\frac{3W}{C^2}\mathcal{Y}^2
+\frac{1}{4\beta^2}}\right].
\label{phlumpo}
\end{eqnarray}
Then, returning to the original (dimensionless) variables and coordinates,
$x$, $y$ and $t$, one may express Eqs.~(\ref{lumpo}) and (\ref{phlumpo}) as follows:
\begin{eqnarray}
\rho_{1}^{(R)}
&\!=\!&-2
\frac{-\epsilon \left[ x-\left(C-\epsilon \frac{V}{8C}\right) t\right]^2
+\epsilon^2 \frac{W}{4C^2}y^2 +\frac{3}{\beta^2}}
{\left[ \epsilon \left[ x-\left(C-\epsilon \frac{V}{8C}\right) t\right]^2
+\epsilon^2 \frac{W}{4C^2}y^2 +\frac{3}{\beta^2} \right]^2},~~~~~~
\label{lumpor} \\[6pt]
\Phi_{0}^{(R)}
&\!=\!& \frac{-\frac{2}{C}\epsilon^{1/2}\left[ x-\left(C-\epsilon \frac{V}{8C}\right) t\right]}
{\epsilon\left[x-\left(C-\epsilon \frac{V}{8C}\right) t\right]^2
+\epsilon^2 \frac{W}{4C^2}y^2 +\frac{3}{\beta^2} }.
\label{lumporph}
\end{eqnarray}
It is clear that upon substituting Eqs.~(\ref{lumpor})-(\ref{lumporph}) into
%It is clear that upon substituting Eqs.~(\ref{lumpo})-(\ref{phlumpo}) into
Eq.~(\ref{psis}), one obtains for the macroscopic wavefunction $\Psi$ an approximate,
vorticity-free, and weakly-localized 2D dark soliton, in the form of a dark lump,
that decays algebraically as $(x^2+y^2)^{1/2} \rightarrow \infty$.
The exciton density $n$, on the other hand, takes the form of an antidark
lump on top of the background density $n_0$, at the location of the dark
polariton soliton, and asymptotes (for $x,y \rightarrow \pm \infty$) to $n_0$.
%Notice that such vorticity-free structures were found to emerge, as a result
%of the slowly developing transverse instability of the 1D dark solitons
%in polariton superfluids in the numerical simulations of Ref.~\cite{ofy}.

%%%%%%%%%%%%%%%%%%%%%%%%%%%%%%%%%%%%%%%%%%%%%%%%%%%%%%%%%%%%%%%%%%%%%%
\begin{figure}[tbp]
	\centering
	\includegraphics[width=0.49\linewidth]{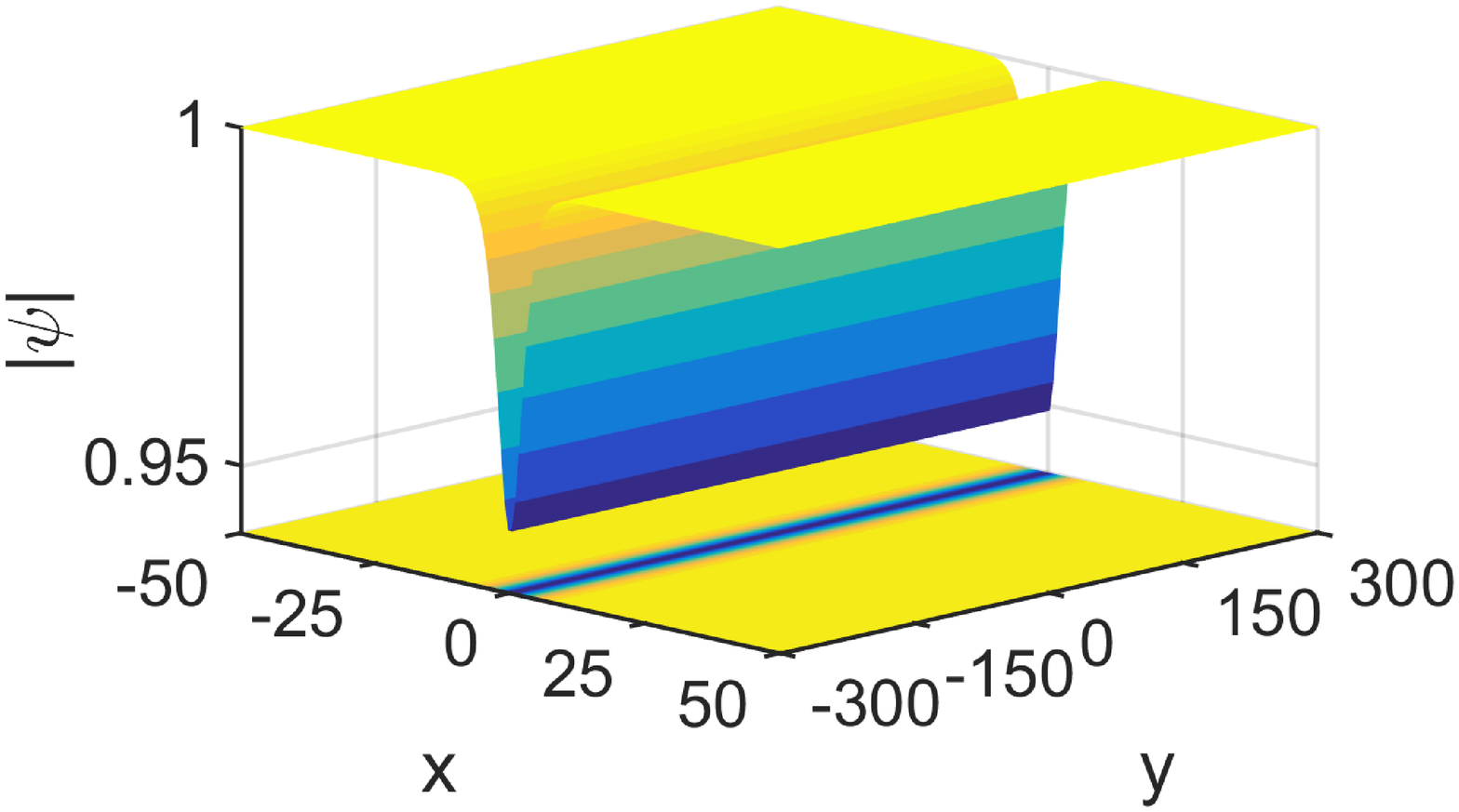}
		\includegraphics[width=0.49\linewidth]{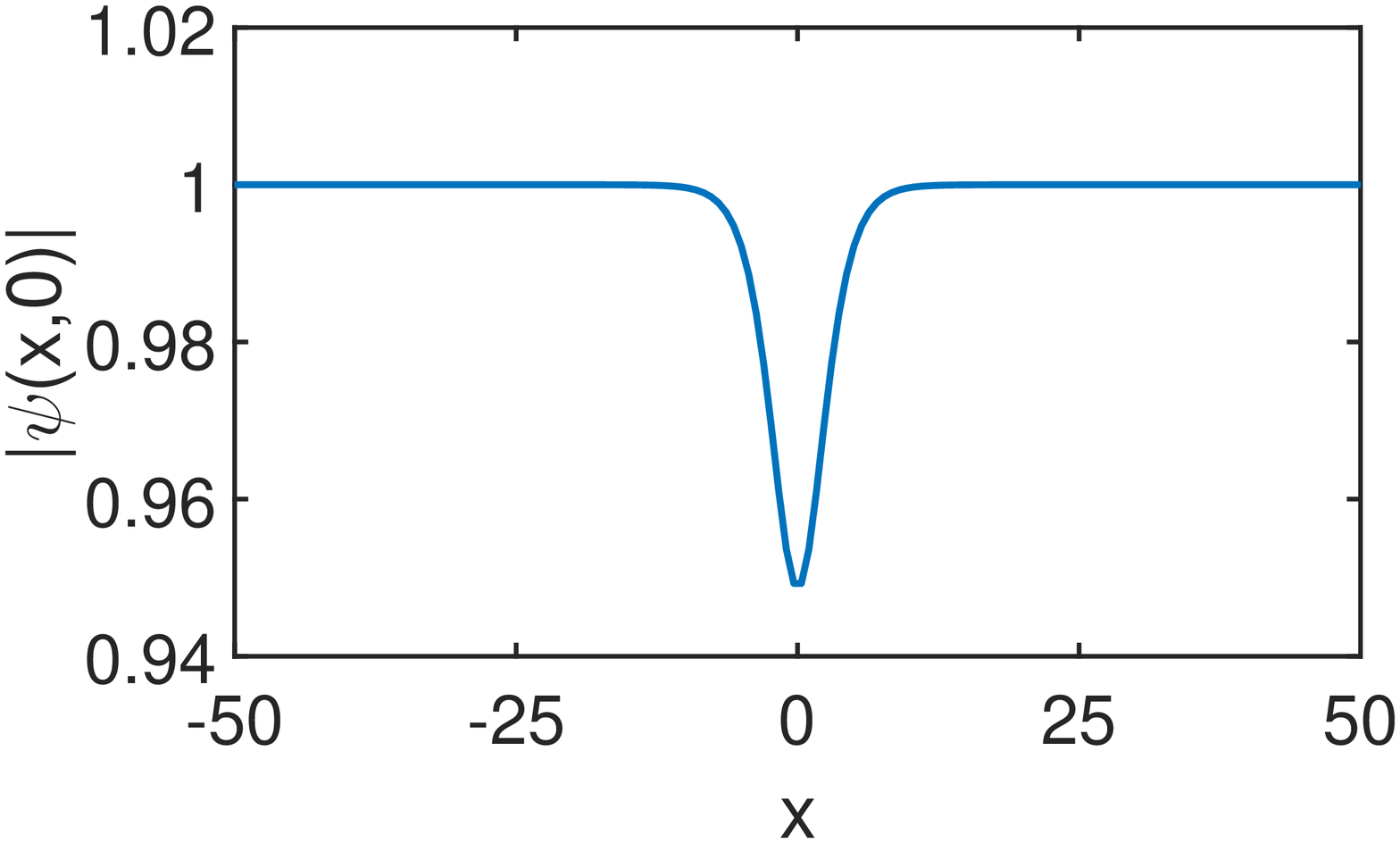}
		\caption{(Color online)
		Three-dimensional plot of the approximate 1D dark stripe soliton (left) and
		the respective wavefunctions' modulus, $|\psi|=|\psi(x,t=0)|$ (right).
		All quantities with tildes are equal to one,
$\gamma_R=g_R=1$, and $\epsilon=0.1$; furthermore, $\kappa=1$.
	}
	\label{fig1}
\end{figure}
%%%%%%%%%%%%%%%%%%%%%%%%%%%%%%%%%%%%%%%%%%%%%%%%%%%%%%%%%%%%%%%%%%%%%%

%%%%%%%%%%%%%%%%%%%%%%%%%%%%%%%%%%%%%%%%%%%%%%%%%%%%%%%%%%%%%%%%%%%%%%
\begin{figure}[tbp]
	\centering
		\includegraphics[width=0.49\linewidth]{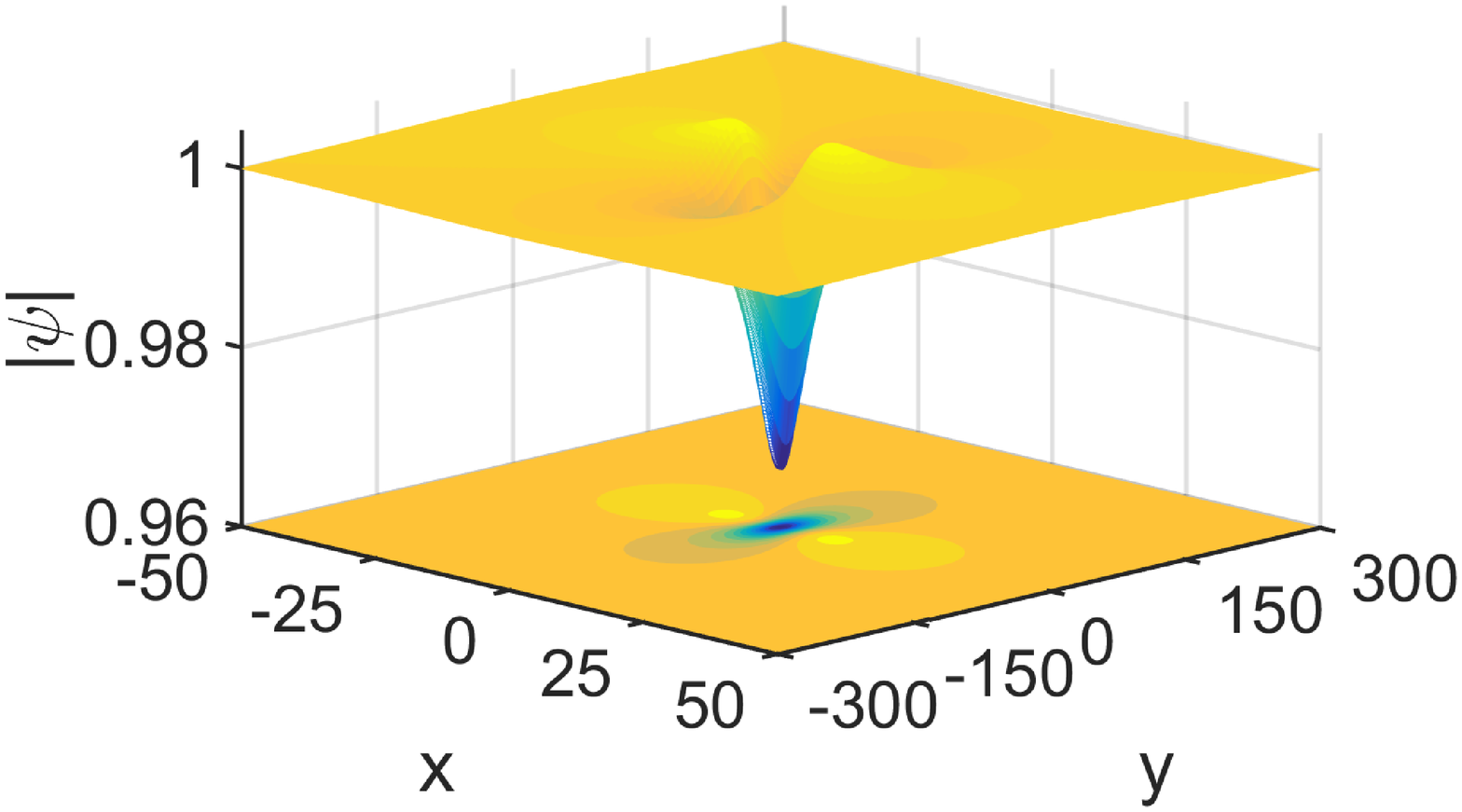}
		\includegraphics[width=0.49\linewidth]{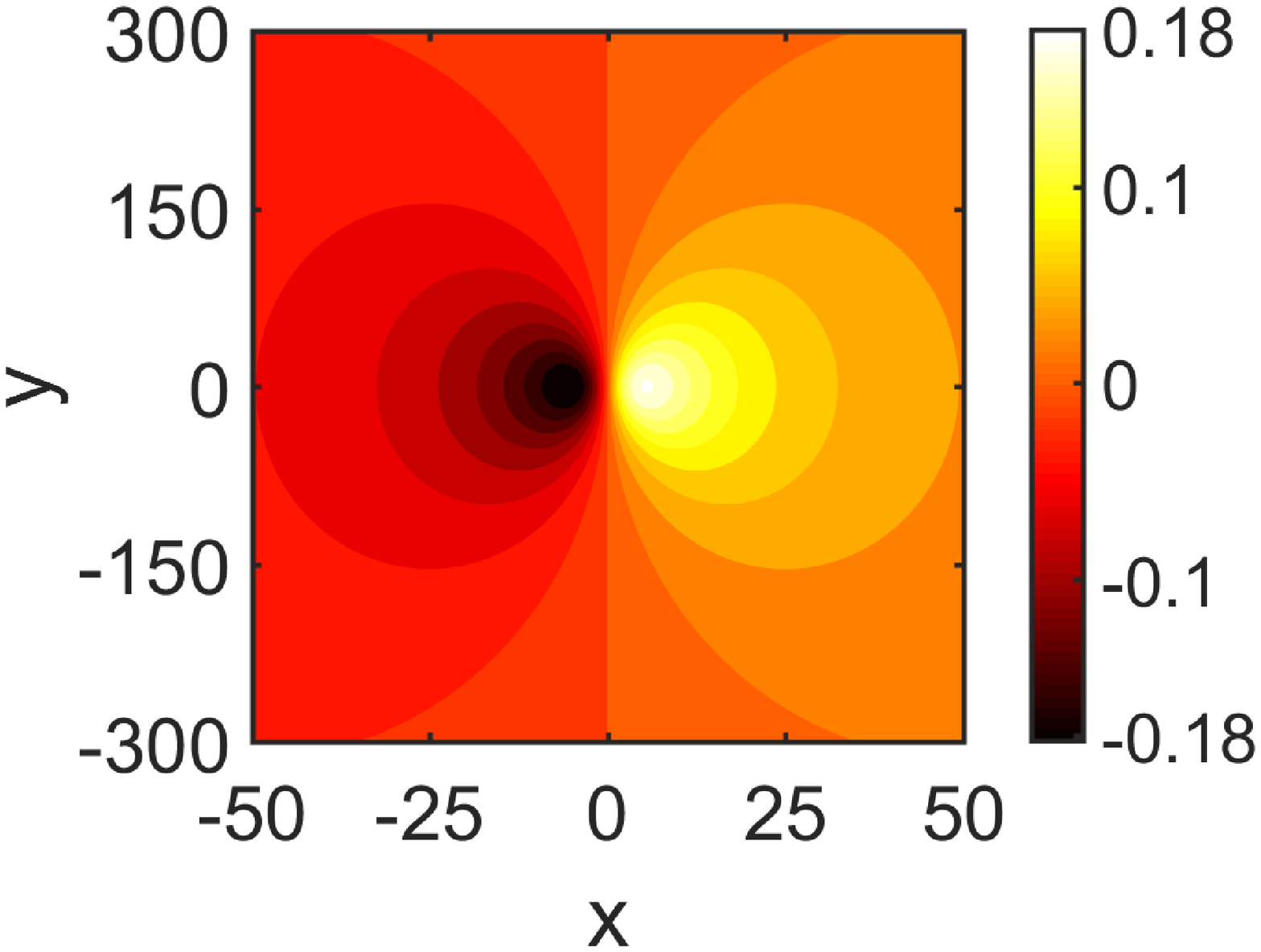} 		
		\includegraphics[width=0.49\linewidth]{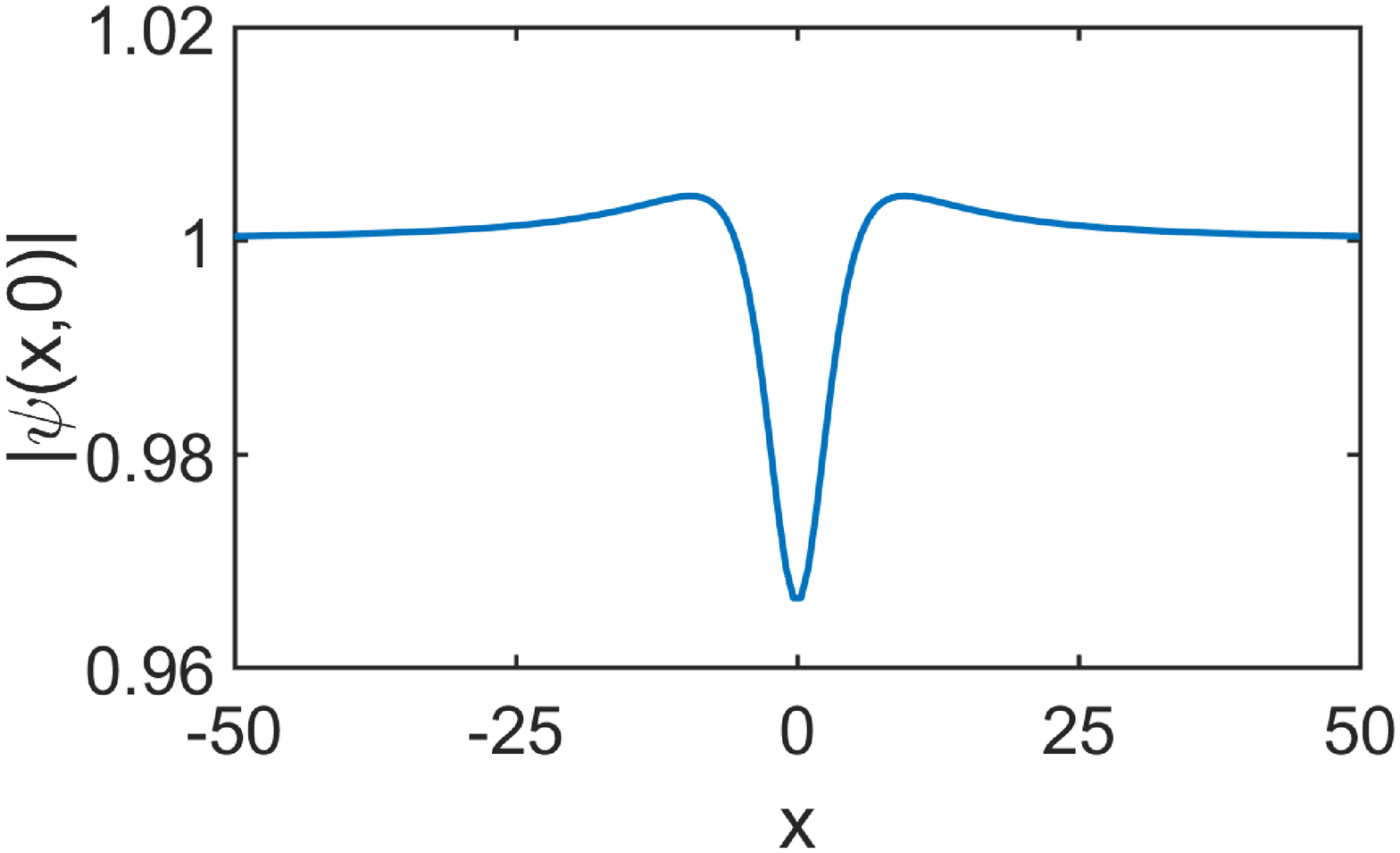}
		\includegraphics[width=0.49\linewidth]{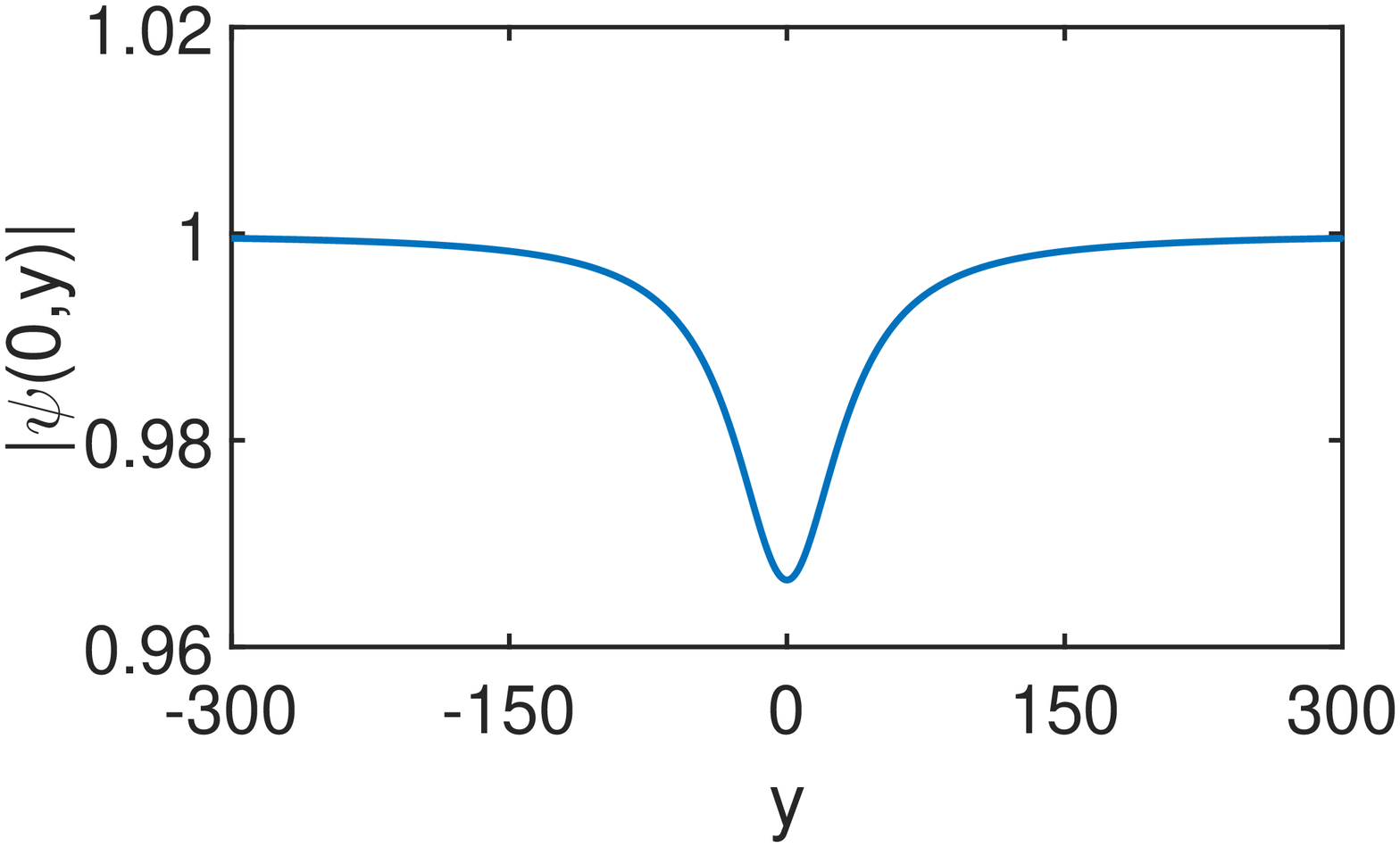}
		\caption{(Color online) Top panels: three-dimensional plot of the approximate
2D dark lump soliton solution (left) and a contour plot depicting its phase profile (right).
Bottom panels: the wavefunction's modulus profiles of the dark lump,
$\psi(x,y=0,t=0)$ (left) and $\psi(x=0,y,t=0)$ (right). All quantities with
tildes are equal to one, $\gamma_R=g_R=1$, and $\epsilon=0.1$; furthermore,
$\beta=1$.
	}
	\label{fig1b}
\end{figure}
%%%%%%%%%%%%%%%%%%%%%%%%%%%%%%%%%%%%%%%%%%%%%%%%%%%%%%%%%%%%%%%%%%%%%%

The form of the approximate soliton solutions, namely of the 1D dark stripe
soliton and the 2D dark lump soliton, is depicted, respectively, 
in Figs.~\ref{fig1} and \ref{fig1b}.
In particular, shown are three-dimensional (3D) plots of the wavefunctions' moduli,
$|\psi|=|\psi(x,t)|$, of the 1D dark soliton stripe (Fig.~\ref{fig1}) and
the 2D dark lump (top panels of Fig.~\ref{fig1b}), at $t=0$; notice that the
plot depicting the lump's phase profile (top right panel of Fig.~\ref{fig1b})
clearly shows that the dark lump is a vorticity-free structure.
For clarity, we also show the spatial profiles of $|\psi|$ for the dark lump,
at $x=0$ and $y=0$ (bottom panels of Fig.~\ref{fig1b}).
Here, we use the following parameter values: all quantities with
tildes are set equal to one, $\gamma_R=g_R=1$,
and $\epsilon=0.1$. In addition, the characteristic parameters of the dark soliton
stripe and the dark lump are respectively chosen to be $\kappa=1$ and $\beta=1$

\subsection{Dissipation-induced soliton dynamics}

Let us now consider the role of the small dissipative perturbation, in the form
of a linear loss term, appearing in the right-hand side of Eq.~(\ref{KP-I}).
%Before proceeding further, it is relevant to note the following. In the context
%of shallow water waves, such KP equations (and also their 1D, KdV, counterparts)
%incorporating a linear loss term have been used in settings involving straits
%of nonuniform water's depth \cite{newell,david}. In such cases, the coefficient
%of the linear loss term (here, $\propto \tilde{\alpha}\tilde{\gamma}_C$)
%is proportional to the (small) gradient of the water's depth.
% PGK: I think you already wrote all this...
%Interestingly, in our case,
%In the present case, the loss coefficient is connected with parameters
%characterizing the open-dissipative nature of the problem, a fact that further establishes   an interesting connection of the polariton superfluids
%problem with the one of shallow water waves.
In both 1D and 2D cases, the evolution of the 1D KdV soliton and the 2D KP-I lump
in the presence of the weak linear loss term has been studied by means
of various techniques.
Let us consider at first the problem of the KdV soliton dynamics in the case $\epsilon \ne 0$,
which has been analyzed in the past by using a perturbed inverse scattering transform (IST)
theory~\cite{karmas,kanew} and asymptotic expansion methods~\cite{ko,kodab} (see also
the review~\cite{kivmal}).
The main result of the analyses reported in these works is that the soliton
has the functional form given in Eq.~(\ref{1ds}), but the parameter $\kappa$
setting the amplitude width and velocity of the soliton, becomes
time-dependent, reflecting the open-dissipative nature of the dynamics.
In particular, in terms of the original time, its evolution is given by the expression:
\begin{equation}
\kappa(t)=\kappa(0)\exp\left(-t/t_{\star}\right),
\label{kapa}
\end{equation}
where $\kappa(0)\equiv \kappa(t=0)$, and the soliton decay rate $t_\star$ is given by:
\begin{equation}
t_\star = \frac{3}{\alpha \gamma_C} t_0
=\frac{3}{\gamma_C} \frac{P_0^{({\rm th})}}{P_0-P_0^{({\rm th})}} ~t_0,
\label{lifetime1}
\end{equation}
where $t_0$ is the characteristic time scale for the system introduced in Sec.~\ref{sec:model}.

On the other hand, the dissipation-induced dynamics of the lump of the KP-I model has
been studied in Ref.~\cite{benilov} by means of the perturbed IST theory.
According to this work, in this case too, in the presence of the weak dissipation
the parameter the parameter $\beta$ characterizing the amplitude width
and velocity of the lump becomes a function of time. In terms of the original time,
its evolution is given by an expression similar to that in Eq.~(\ref{lifetime1}), namely:
\begin{equation}
\beta(t)=\beta(0)\exp\left(-t/T_{\star}\right),
\label{beta}
\end{equation}
where $\beta(0)\equiv \beta(t=0)$, with the lump decay rate $T_\star$ given by:
\begin{equation}
T_\star = \frac{1}{\alpha \gamma_C} t_0
=\frac{1}{\gamma_C} \frac{P_0^{({\rm th})}}{P_0-P_0^{({\rm th})}} ~t_0.
\label{lifetimel}
\end{equation}
It is observed that, in the weak pumping regime under consideration,
both the soliton's and the lump's decay rates
depend on the decay rate $\gamma_C$ of the polariton condensate, as well as
the relative deviation $\alpha$ of the uniform pumping $P_0$
from the threshold value $P_0^{({\rm th})}$.
Importantly, it turns out that
\begin{equation}
T_\star = \frac{1}{3}\, t_\star,
\end{equation}
a fact that indicates that the soliton decays faster than the lump.
Furthermore, it is relevant to note that the soliton stripe,
in addition to the aforementioned decay,
is also subject to transverse instabilities,
as we will discussion in our numerical results below.
Thus, chiefly,
the weakly localized 2D dark solitonic structure constitutes the
2D (non-vortical)
coherent structure that has the best chance to be observed in
realistic physical experiments.

\subsection{Numerical results}

Let us now test the validity of our analytical considerations above by contrasting them
against direct numerical simulations of the original system of Eqs.~(\ref{Psin})-(\ref{nRn}).
In particular, for this comparison, we will study:
\begin{itemize}
\item[(i)] the existence and dissipative dynamics of dark lumps, and
\item[(ii)] the spontaneous generation of
  coherent structures resulting from the snaking instability
of dark soliton stripes.
\end{itemize}

For simplicity, in our simulations, all quantities with tildes
have been set equal to unity, namely $\tilde{\alpha}=\tilde{\gamma}_C=\tilde{R}=1$,
as well as $\gamma_R=g_R=1$. Thus, parameters of 
the open dissipative GP model, Eqs.~(\ref{Psin})-(\ref{nRn}), as well as our initial data
only depend on the small parameter $\epsilon$. For the initial conditions, we use the analytical
form of the dark soliton stripe and dark lump, i.e., Eqs.~(\ref{psis})-(\ref{ns}),
with $\rho_{1}^{(R)}$ and $\Phi$ given by Eqs.~(\ref{1ds}) and (\ref{1dsph}) for the stripe
or by Eqs.~(\ref{lumpo}) and (\ref{phlumpo}) for the lump.

%%%%%%%%%%%%%%%%%%%%%%%%%%%%%%%%%%%%%%%%%%%%%%%%%%%%%%%%%%%%%%%%%%%%%%
\begin{figure}[tbp]
	\centering
	\includegraphics[width=0.49\linewidth]{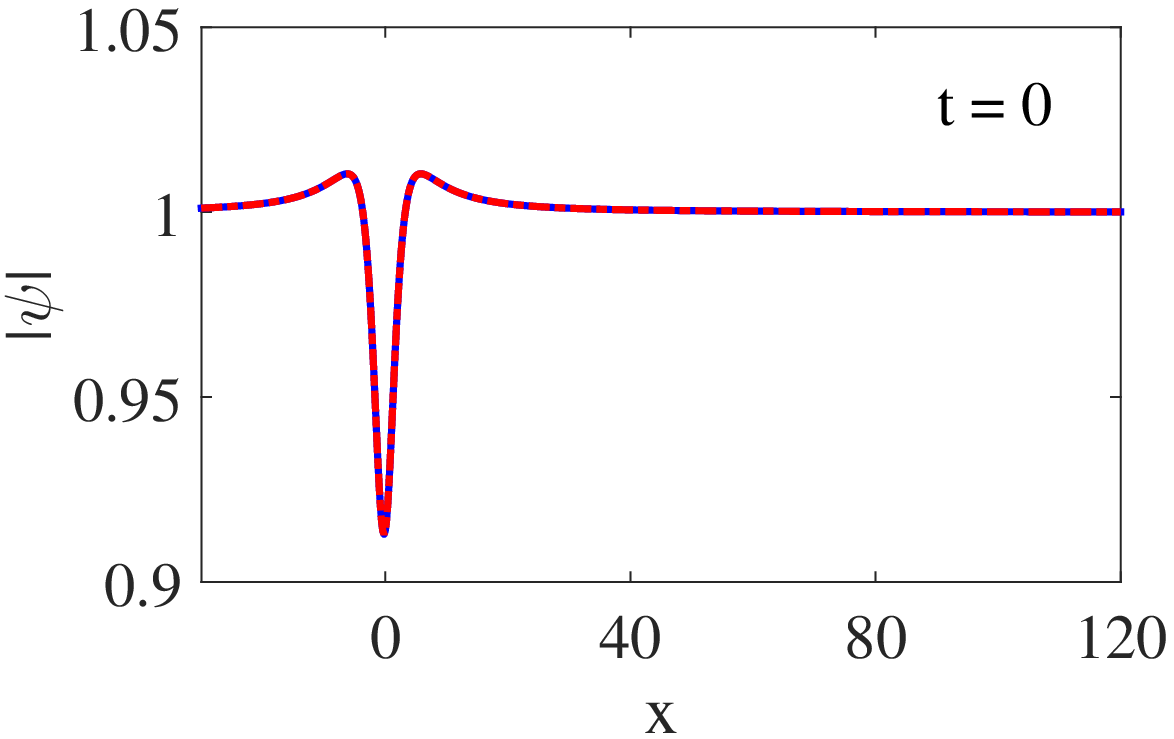}
	\includegraphics[width=0.49\linewidth]{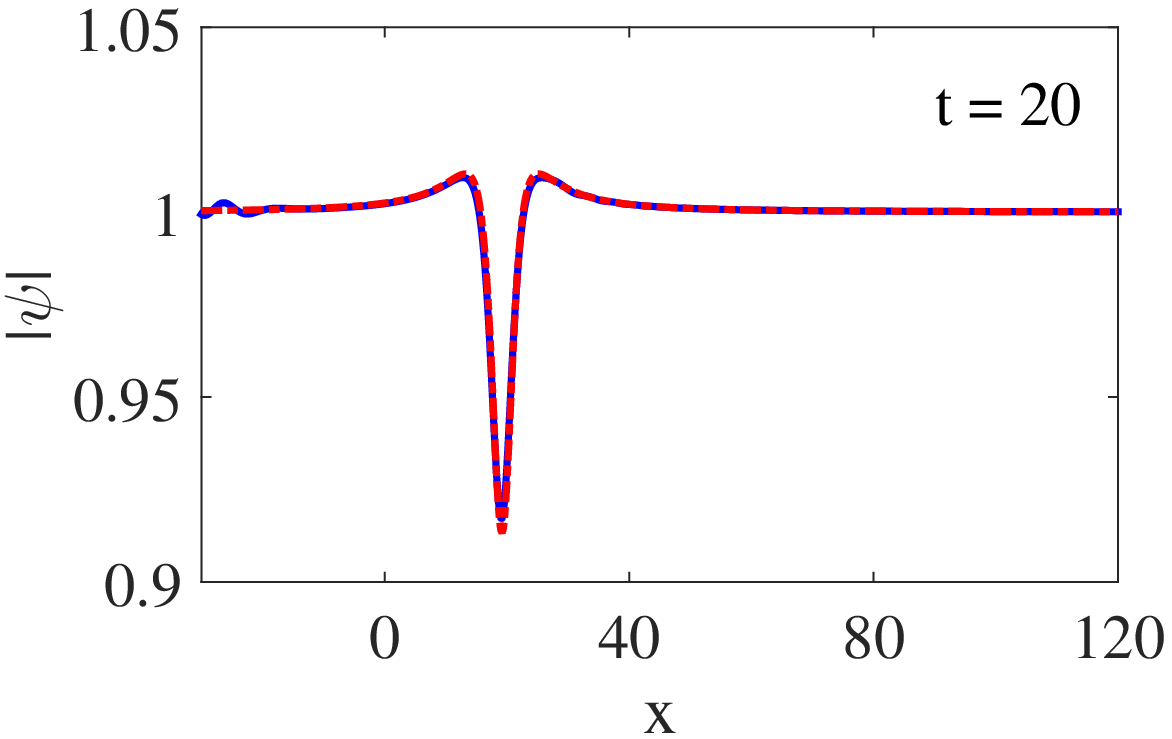}
	\includegraphics[width=0.49\linewidth]{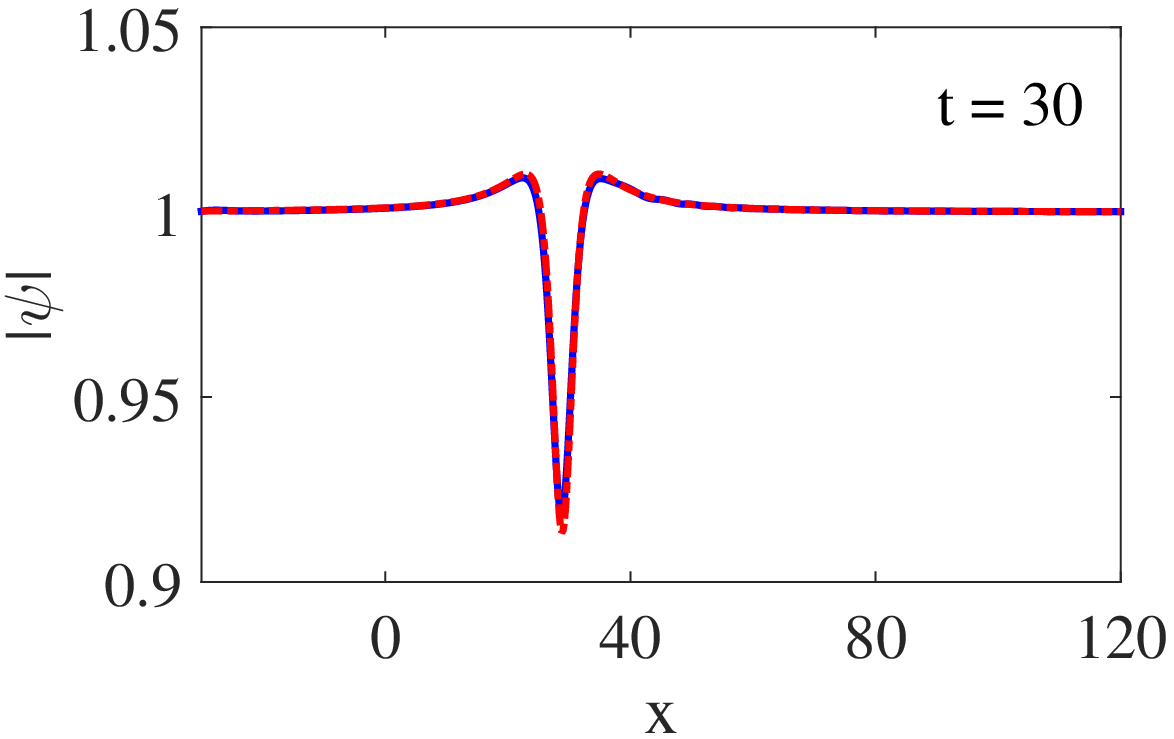}
	\includegraphics[width=0.49\linewidth]{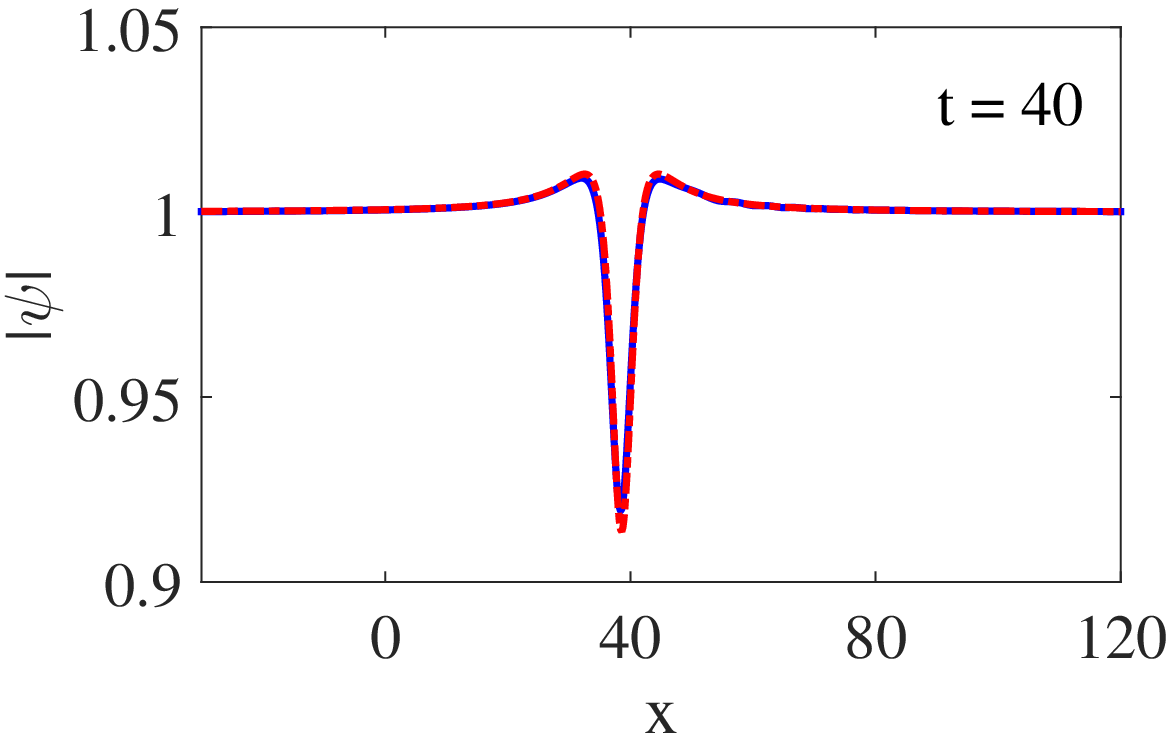}
	\includegraphics[width=0.49\linewidth]{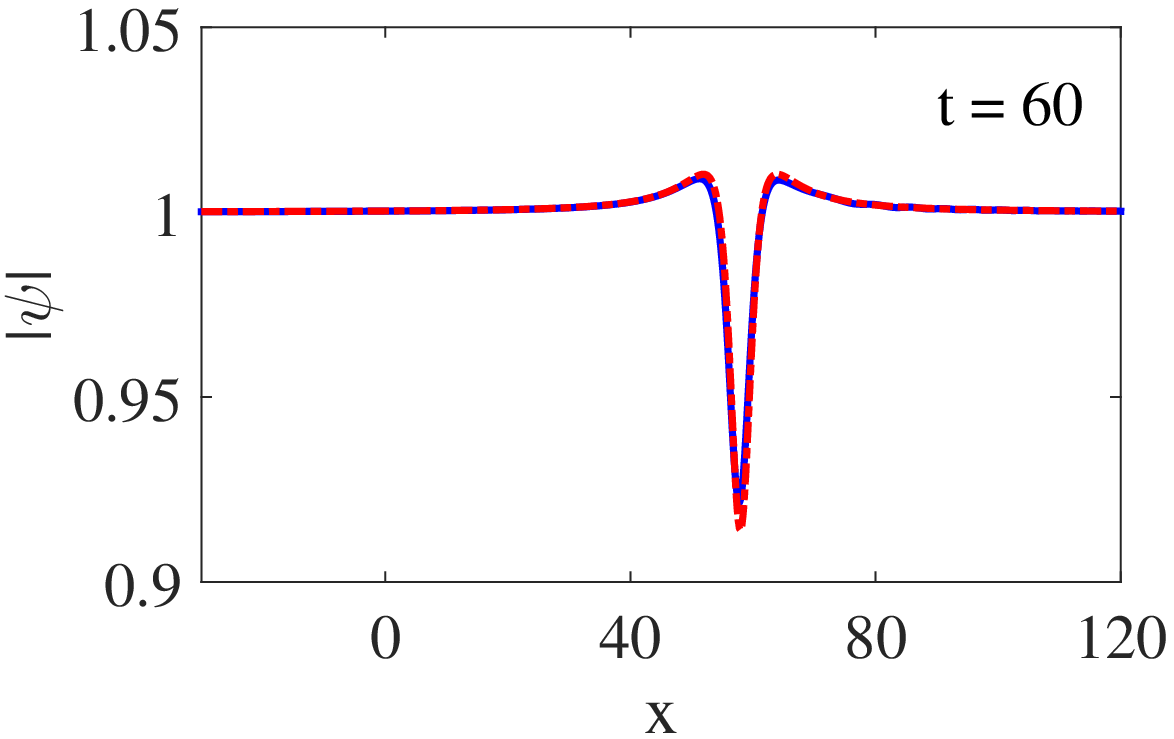}
	\includegraphics[width=0.49\linewidth]{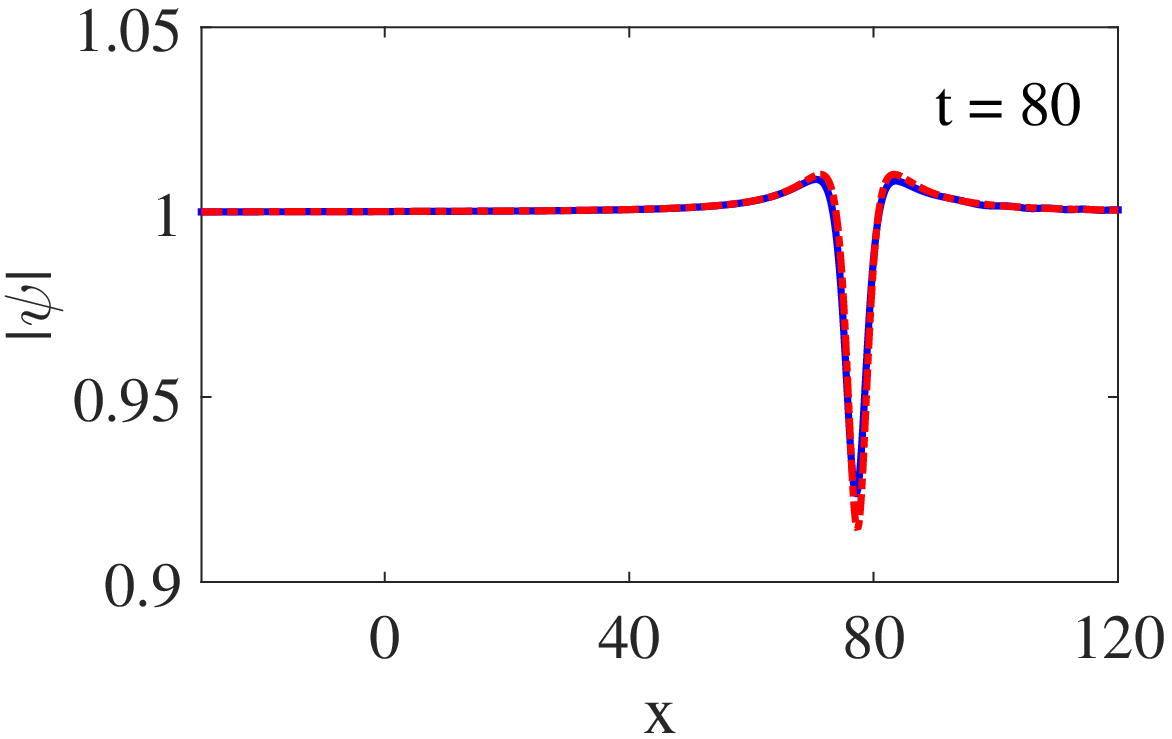}
\caption{(Color online) 
Evolution of the dark lump soliton.
Solid (dashed) lines depict the numerical (analytical)
wavefunction's modulus $\psi(x,y,t)$ for $y=0$.
All quantities with tildes are equal to one, 
$\gamma_R=g_R=1$, and $\epsilon=0.01$; furthermore, $\beta(0)=5$.
	}
	\label{fig2}
\end{figure}
%%%%%%%%%%%%%%%%%%%%%%%%%%%%%%%%%%%%%%%%%%%%%%%%%%%%%%%%%%%%%%%%%%%%%%

%%%%%%%%%%%%%%%%%%%%%%%%%%%%%%%%%%%%%%%%%%%%%%%%%%%%%%%%%%%%%%%%%%%%%%
\begin{figure}[tbp]
	\centering
    \includegraphics[scale=0.3]{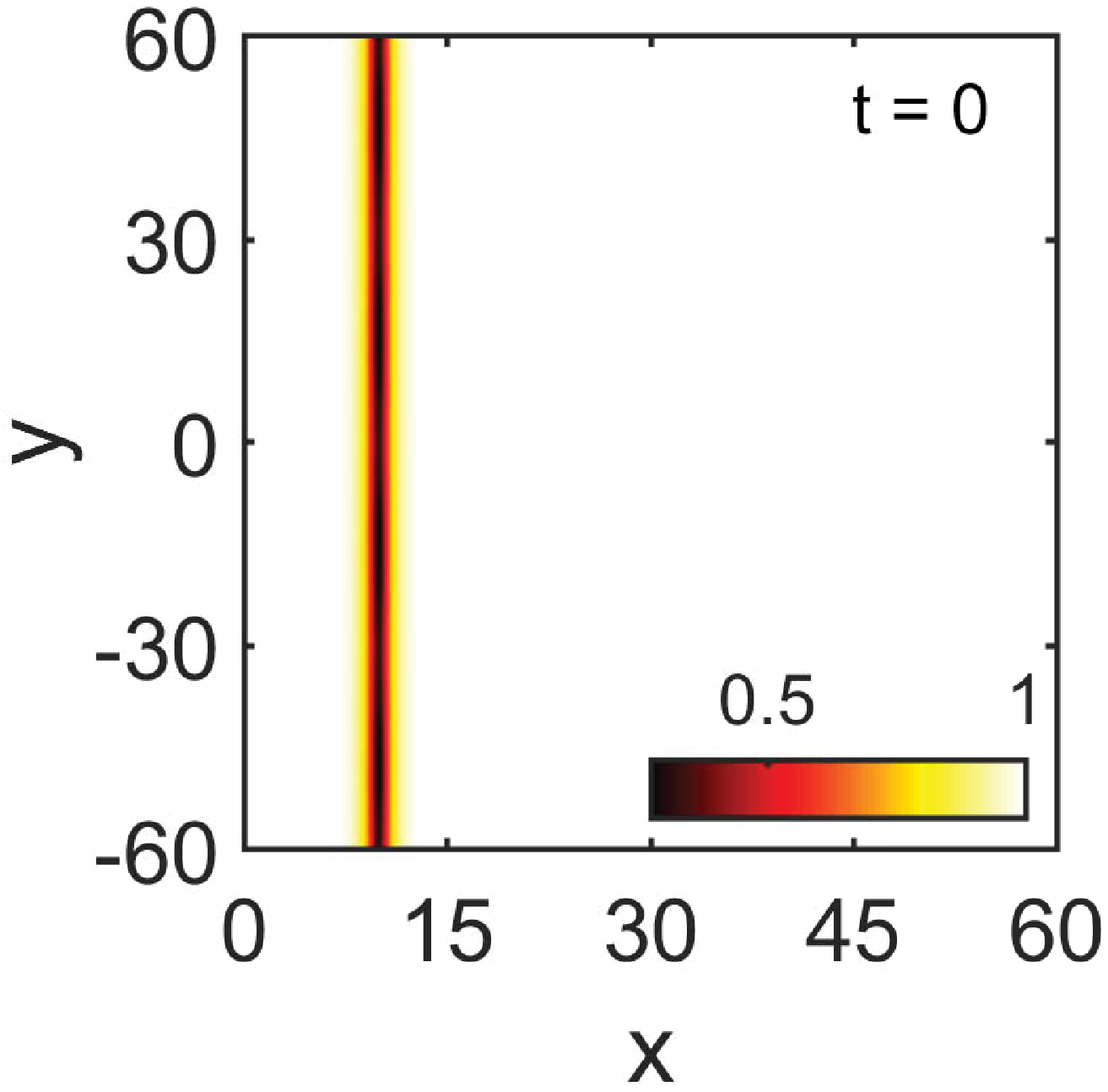}
	\includegraphics[scale=0.3]{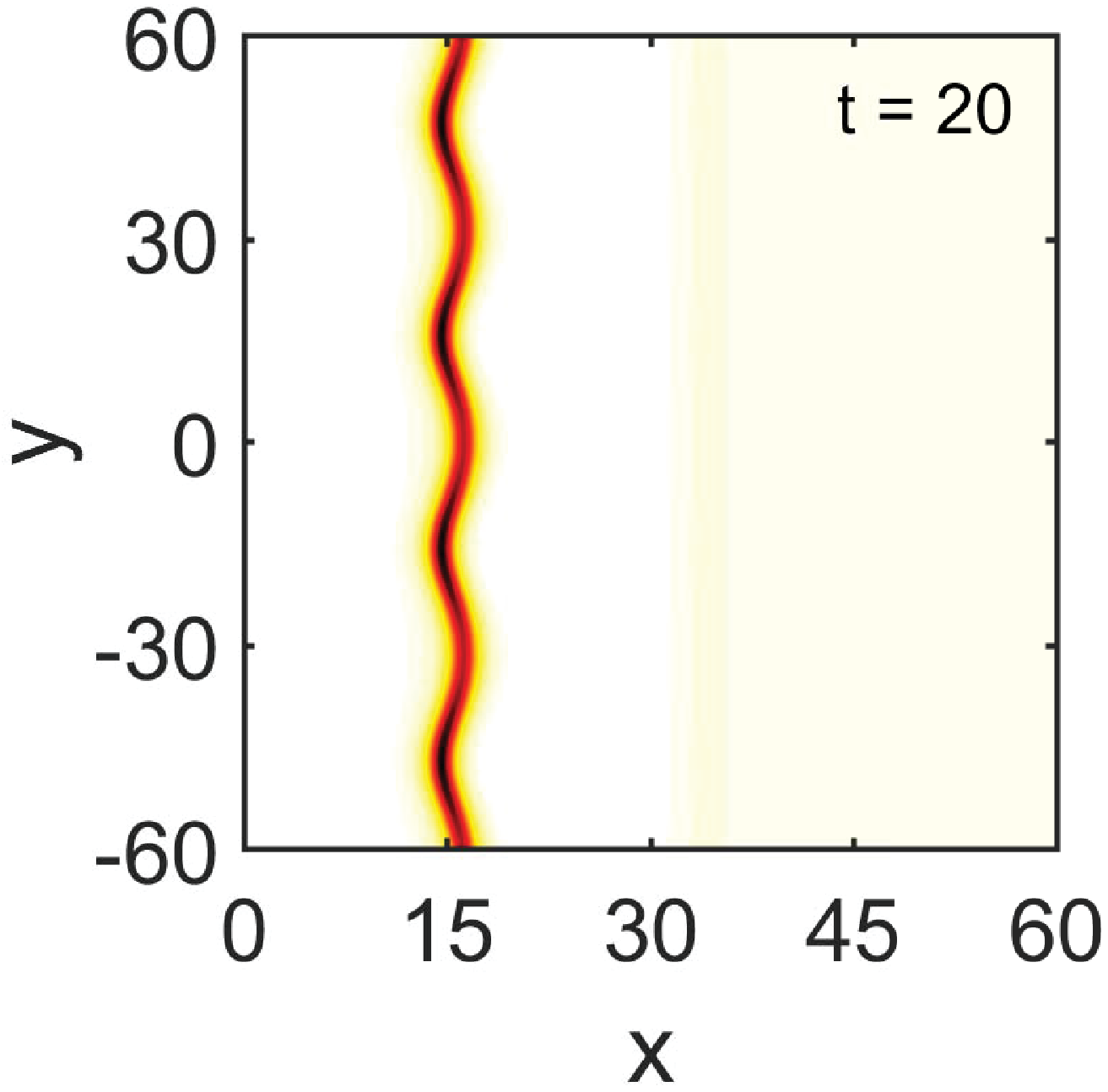}
	\includegraphics[scale=0.3]{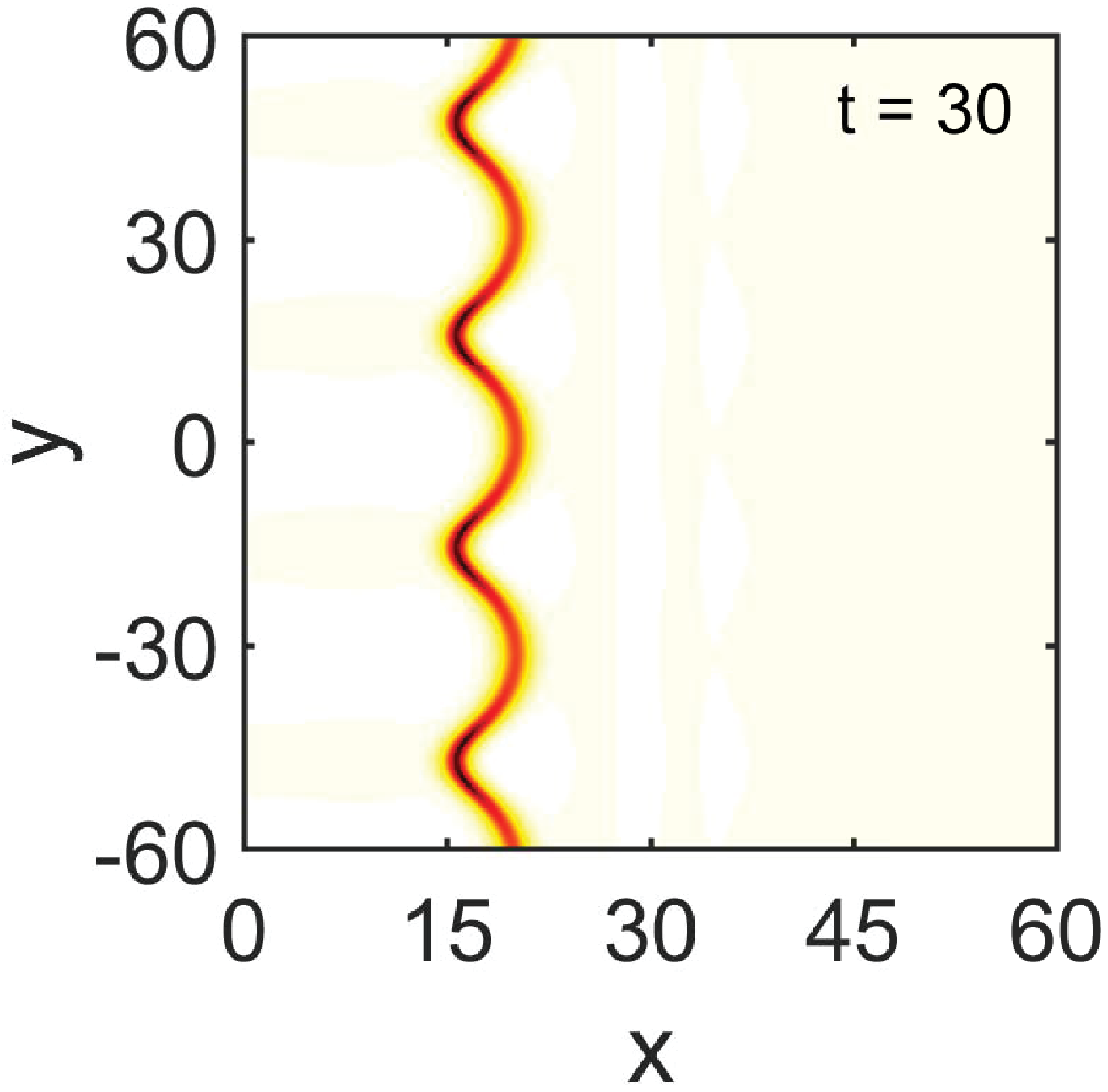}
	\includegraphics[scale=0.3]{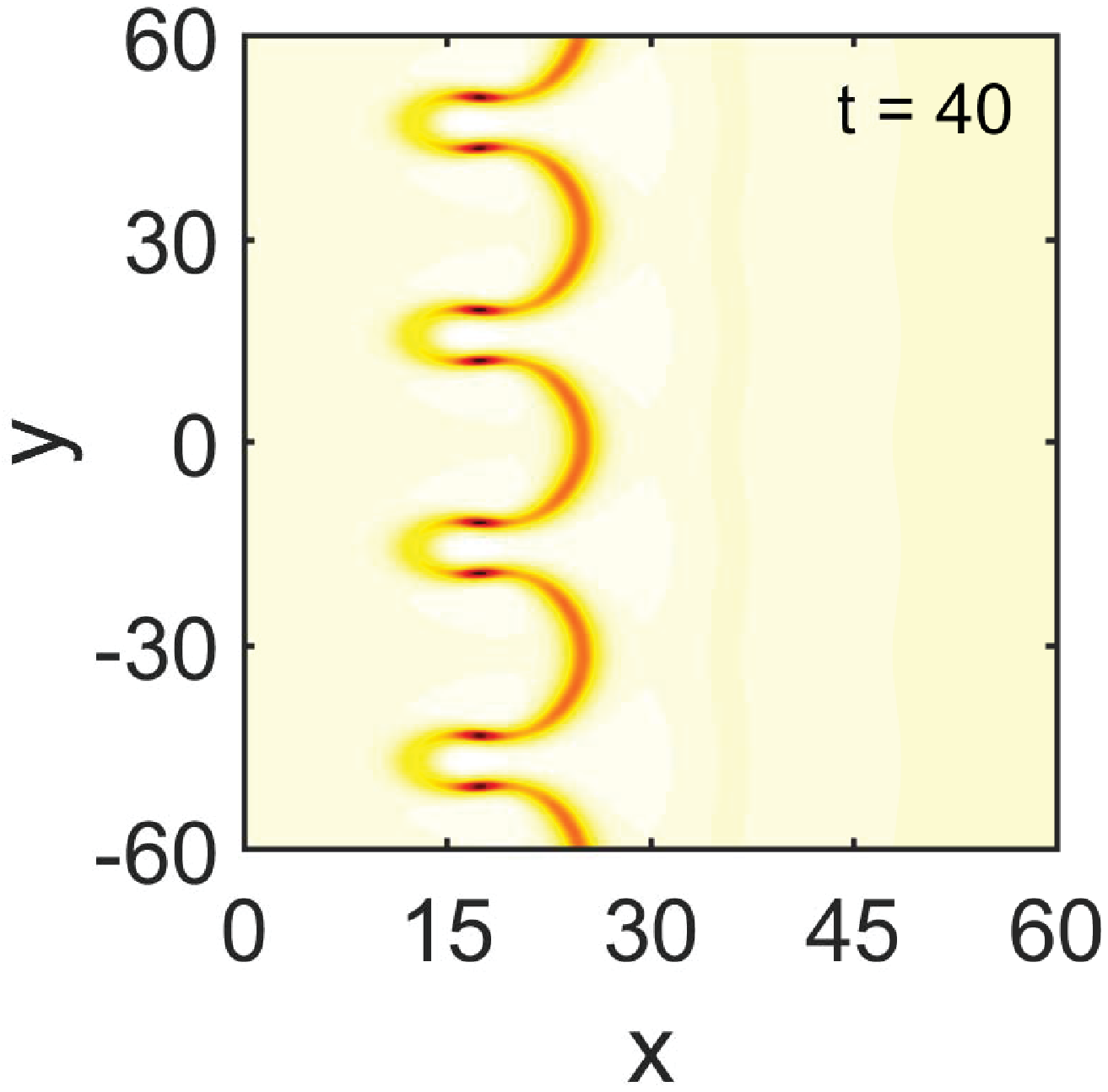}
	\includegraphics[scale=0.3]{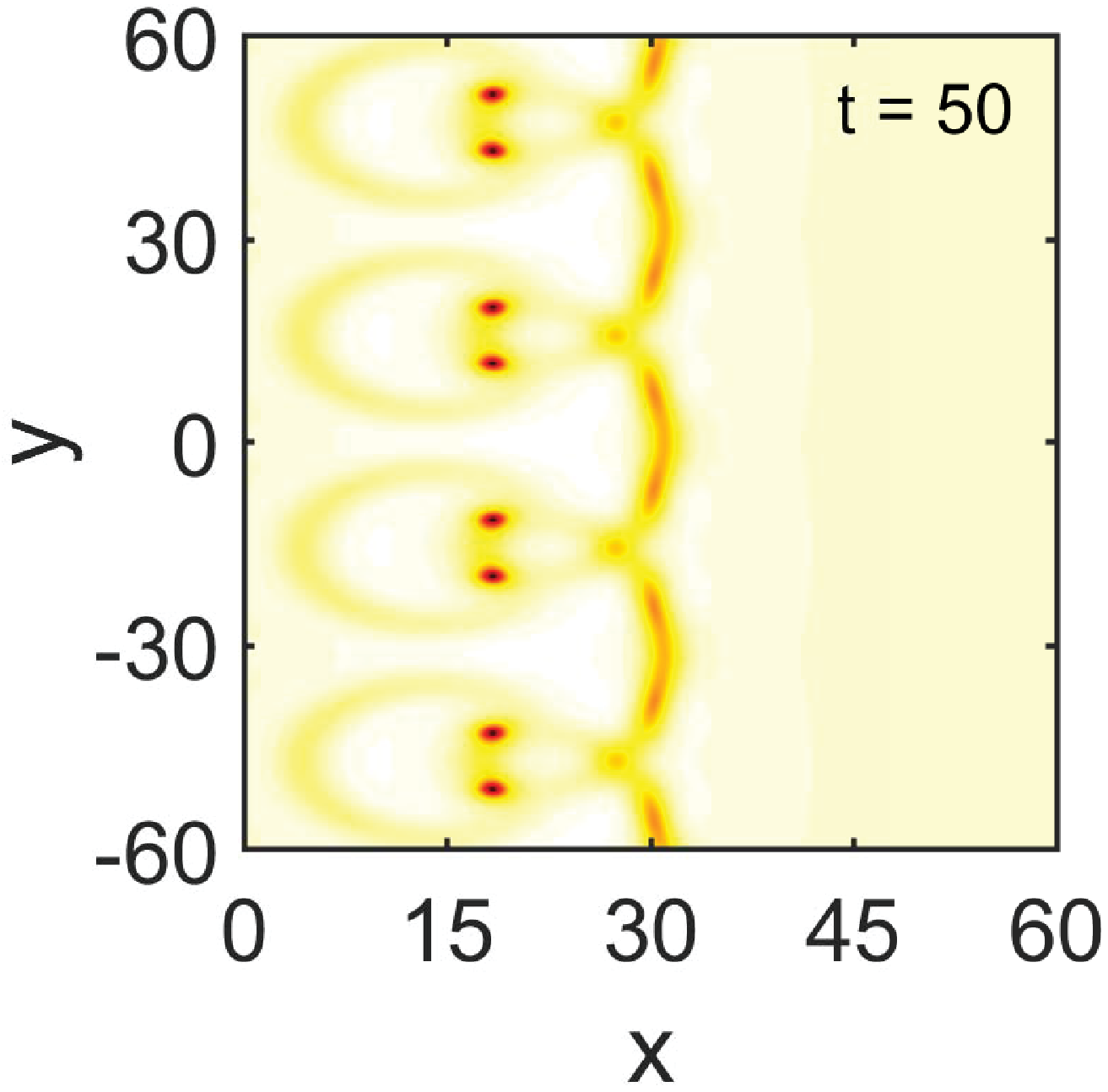}
	\includegraphics[scale=0.3]{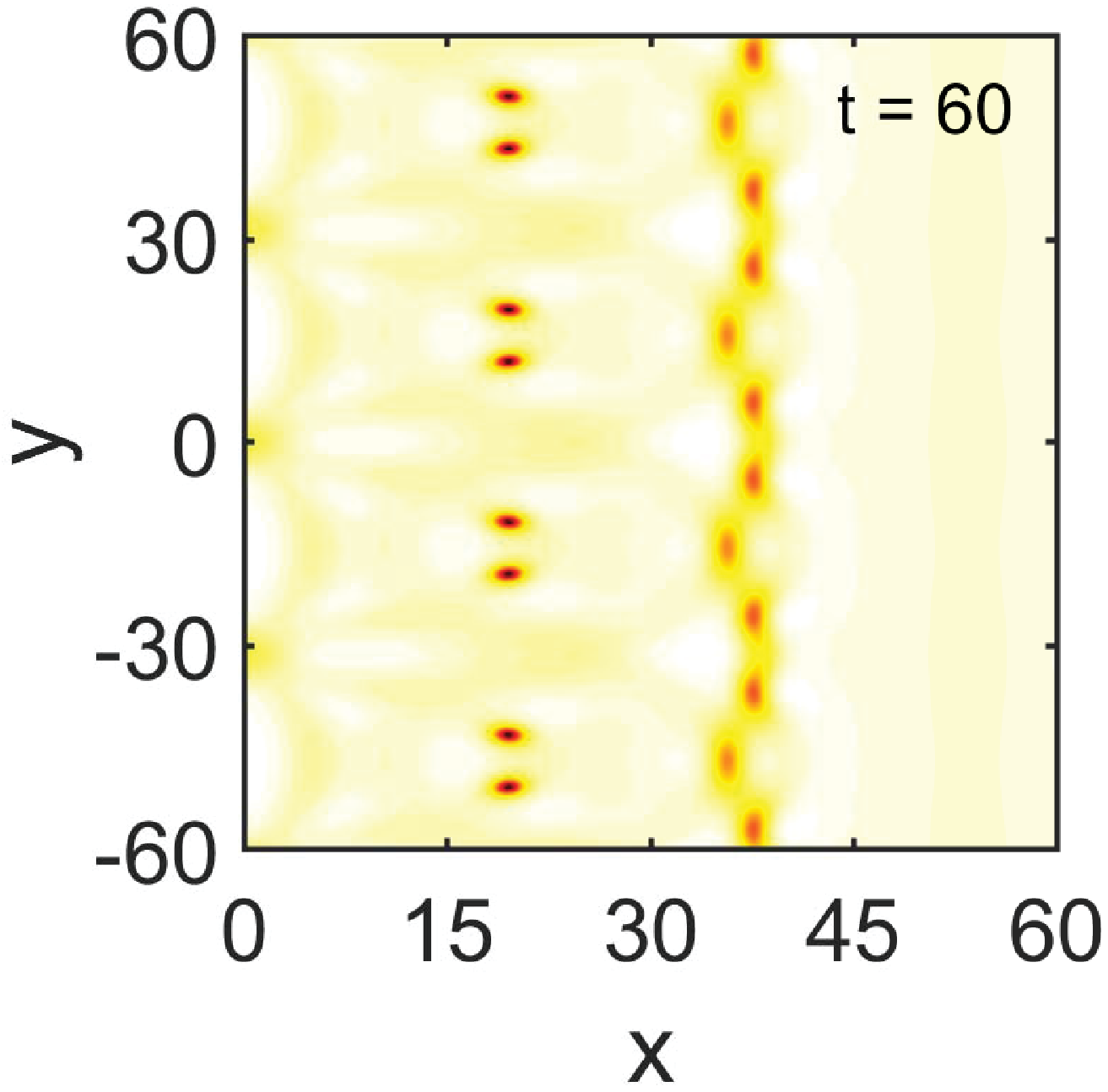}
	\caption{(Color online) 
Evolution of a relatively deep dark soliton stripe undergoing snaking instability.
The panels depict the wavefunction's modulus at the indicated times.
Notice that after strong undulation, the stripe decays into vortex pairs,
which are formed at approximately $t=50$. 
A zoom showing the lowest vortex pair is depicted in Fig.~\ref{fig3b}.
All quantities with tildes
are equal to one, $\gamma_R=g_R=1$, and $\epsilon=0.1$; furthermore, here, 
$\kappa=3+0.3\cos(0.2y)$.
	}
	\label{fig3}
\end{figure}
%%%%%%%%%%%%%%%%%%%%%%%%%%%%%%%%%%%%%%%%%%%%%%%%%%%%%%%%%%%%%%%%%%%%%%

First, we study the existence and dissipative dynamics of dark lumps
(for a relevant study for the dark solitons in the 1D setting see Ref.~\cite{wepla}).
Since our approach relies on a perturbation method, it is expected that
the agreement between analytical and numerical results will be better
for relatively smaller values of $\epsilon$; for this reason, we choose
$\epsilon=0.01$. Figure~\ref{fig2} shows snapshots of the dark lump modulus
$|\psi(x,y,t)|$ as a function of $x$ (i.e., along the
propagation direction), for different time instants, up to $t=80$.
We have obtained similar results (in terms of the quality of the
agreement with the theoretical prediction) along the $y$-direction
(results not shown here).
It is observed that dark lump solitons do exist and they follow dissipative dynamics which
is well described by the analytical predictions ---compare the numerical (solid lines)
and analytical (dashed lines) profiles of the lump modulus. For this simulation, the
relative maximum error in the estimation of the dark lump's minimum at $t=80$
is less than $3\%$. We have checked that even for values of $\epsilon$ about an
order of magnitude larger, the error is approximately $15\%$ (at the same time $t=80$),
while the qualitative characteristics of the lump's evolution are
similar to those shown in Fig.~\ref{fig2}. In any case, we have found that
our analytical approach tends to underestimate the actual dissipation of the 
dark lump solitons.

Next, having checked the existence and dissipation-induced dynamics of
the dark lumps, we now proceed to study the evolution of the
dark line (i.e., stripe) solitons. Here, we focus on
their spontaneous breakup resulting from the transverse instability.
As discussed in the previous section, in the 2D setting,
line solitons of KP-I are unstable and decay into lumps \cite{infeld}. In the
context of the defocusing NLS, the snaking instability of dark soliton stripes
results in their decay into vortices \cite{kuz1}; this effect was 
studied in detail also in the context of polariton superfluids \cite{ofy}.

The connecting link between these two pictures is the formal asymptotic reduction
of the open dissipative GP model to the KP-I equation: based on previous results
referring to shallow solitons of the 2D NLS \cite{smirnov1}, we expect that
sufficiently deep stripe dark solitons, which are beyond the analytical description
of Eqs.~(\ref{psis})-(\ref{ns}) will decay into vortices; on the contrary, sufficiently
shallow stripe dark solitons, described by Eqs.~(\ref{psis})-(\ref{ns}),
with $\rho_{1}^{(R)}$ and $\Phi$ as given by Eqs.~(\ref{1ds}) and (\ref{1dsph}),
will develop undulations in 2D and eventually decay into 2D vorticity-free
structures resembling the dark lump solitons.
% PGK: I am not sure I subscribe to this...(nor to the fact
% that this is what the figure shows...). I would amend the text
% towards a more neutral description...

%%%%%%%%%%%%%%%%%%%%%%%%%%%%%%%%%%%%%%%%%%%%%%%%%%%%%%%%%%%%%%%%%%%%%%
\begin{figure}[tbp]
	\centering
	\includegraphics[scale=0.25]{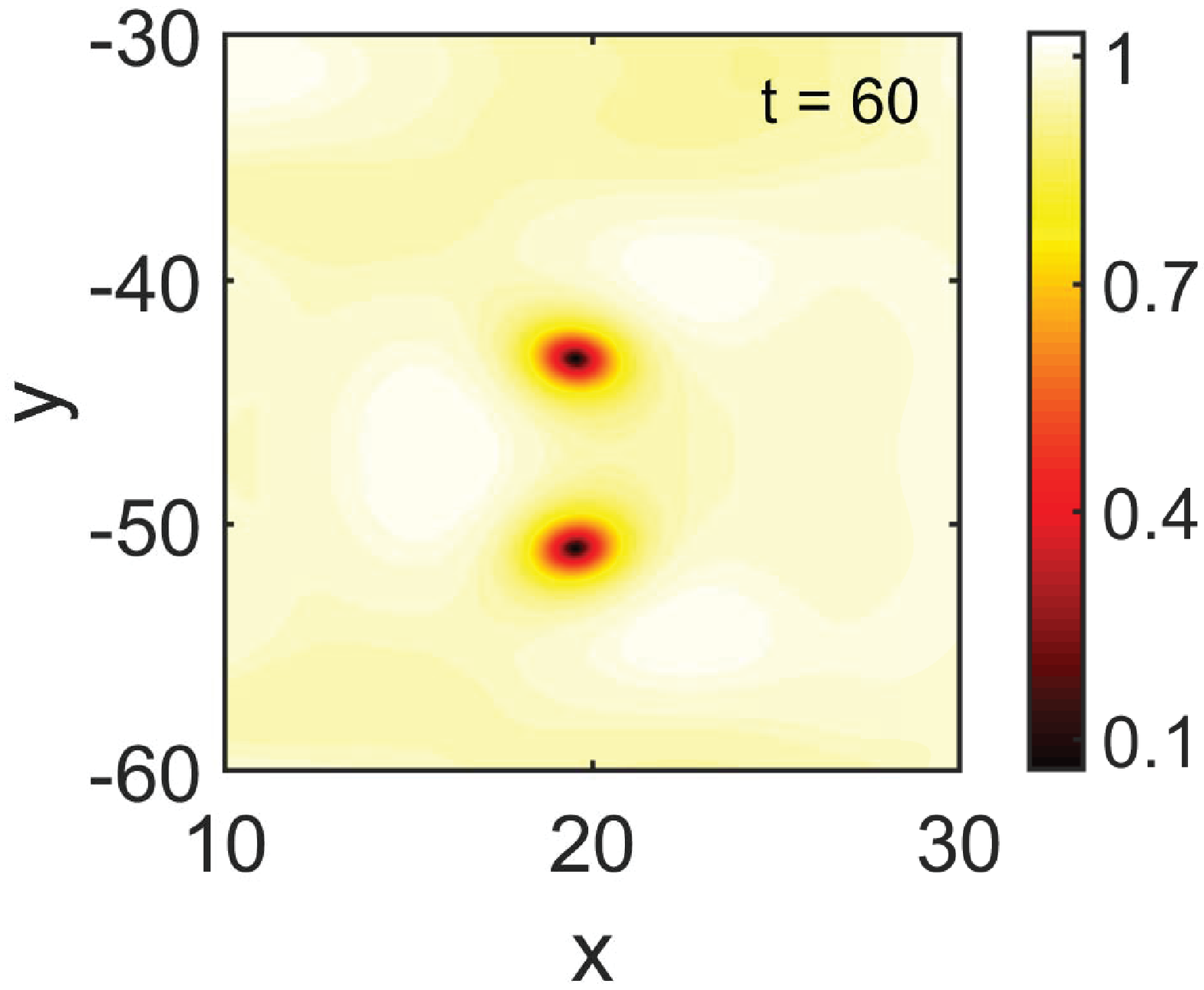}
	\includegraphics[scale=0.25]{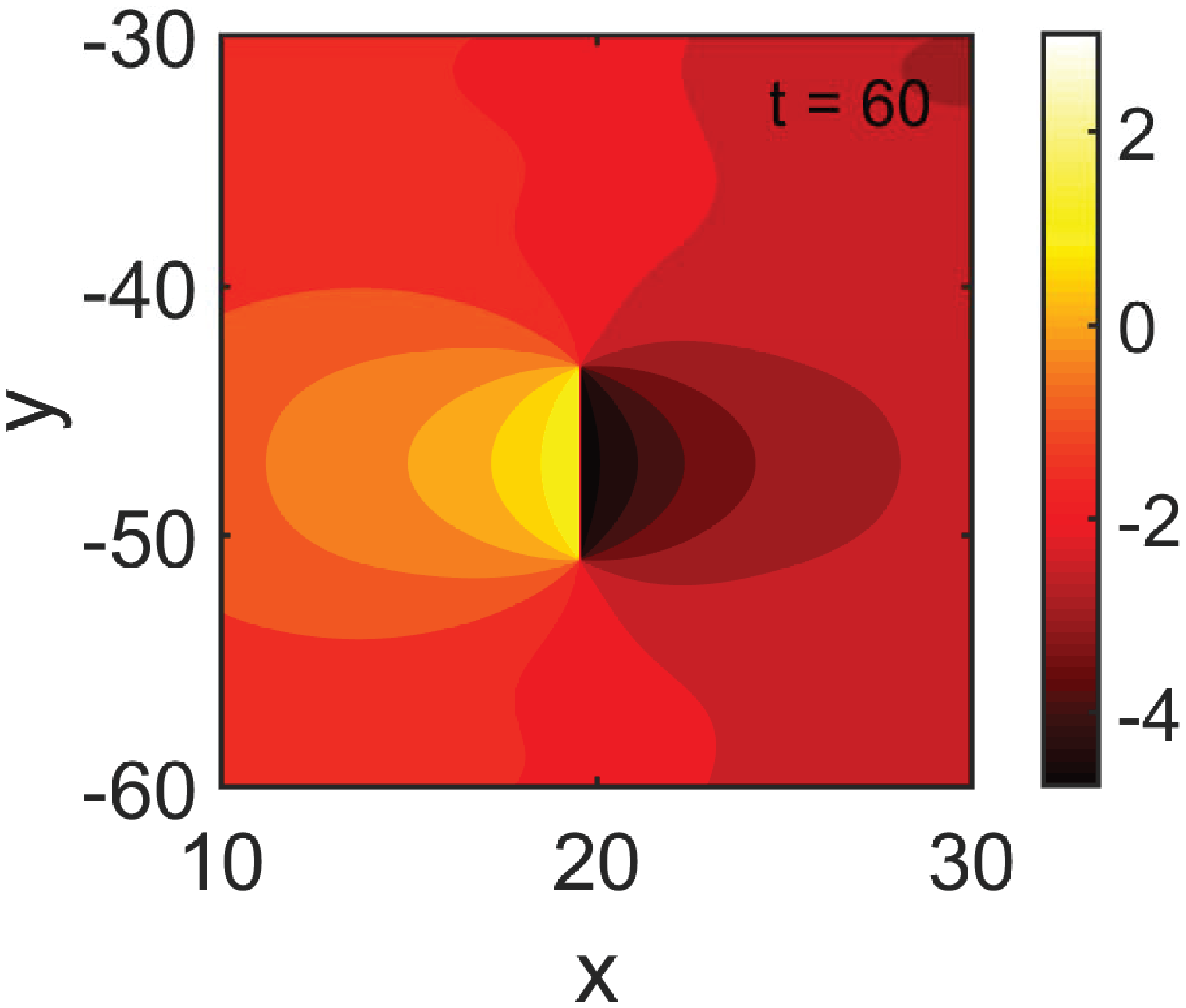}
	\includegraphics[width=0.49\linewidth]{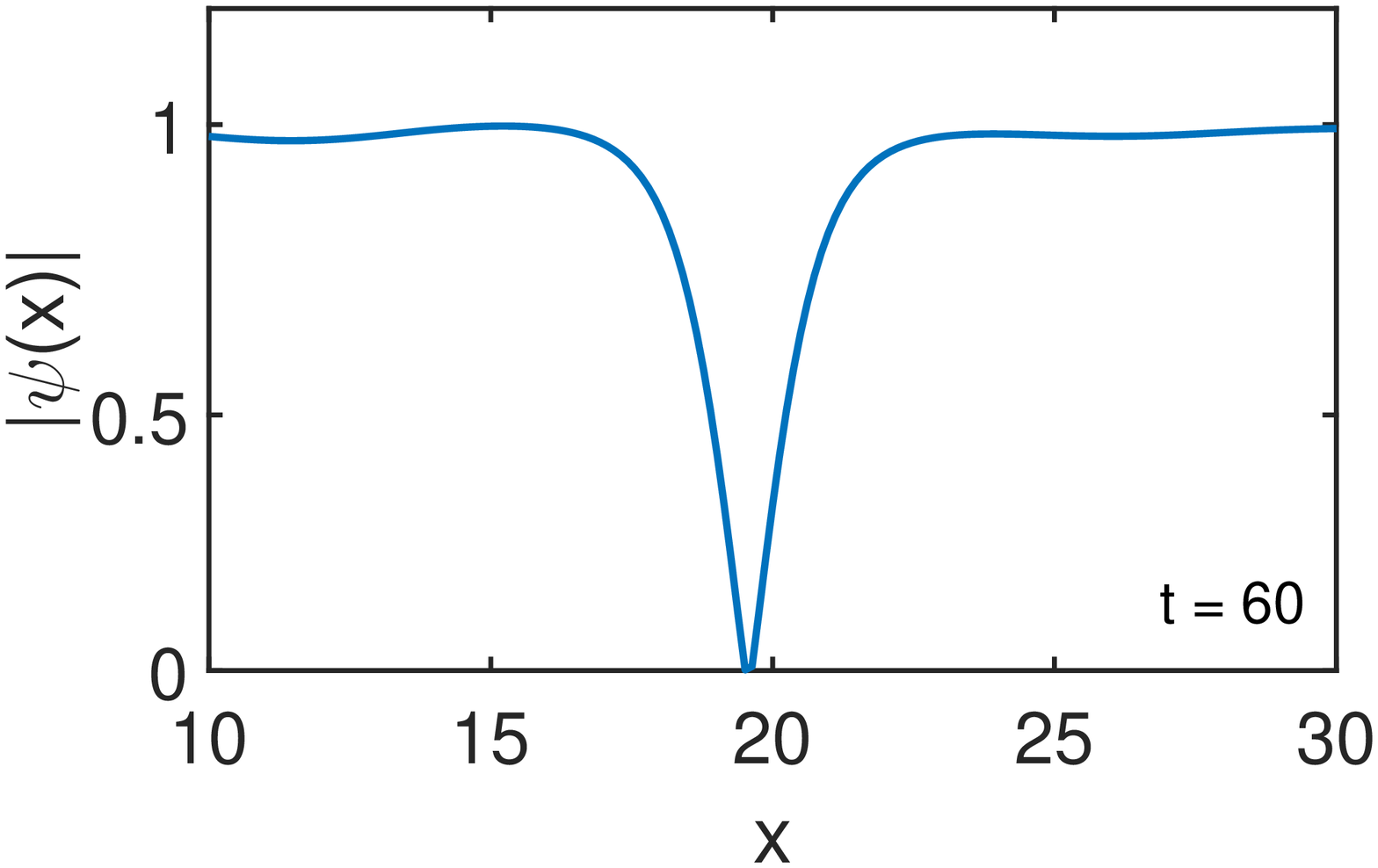}
	\includegraphics[width=0.49\linewidth]{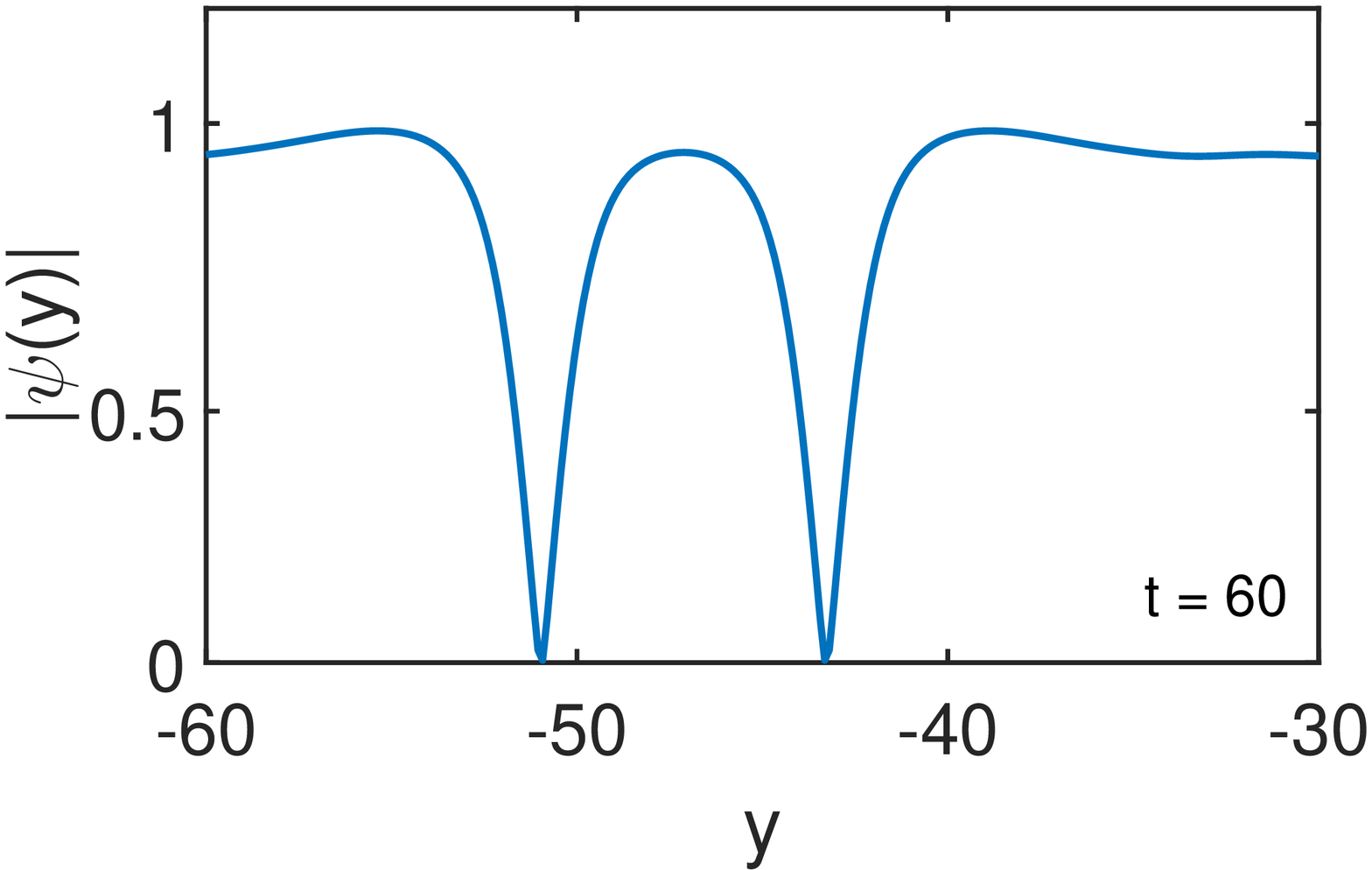}
		\caption{(Color online)	
A zoom depicting the lowest vortex pair 
shown in Fig.~\ref{fig3} at $t=60$. The top panels correspond to contour plots 
of the density (left) and the phase (right); the latter, clearly reveals
the phase profile of a vortex-antivortex pair. The bottom panels depict the $x$- (left) 
and $y$- (right) profiles of wavefunction's modulus of the vortex pair.
	}
	\label{fig3b}
\end{figure}
%%%%%%%%%%%%%%%%%%%%%%%%%%%%%%%%%%%%%%%%%%%%%%%%%%%%%%%%%%%%%%%%%%%%%%

%%%%%%%%%%%%%%%%%%%%%%%%%%%%%%%%%%%%%%%%%%%%%%%%%%%%%%%%%%%%%%%%%%%%%%
\begin{figure}[tbp]
	\centering
	\includegraphics[scale=0.3]{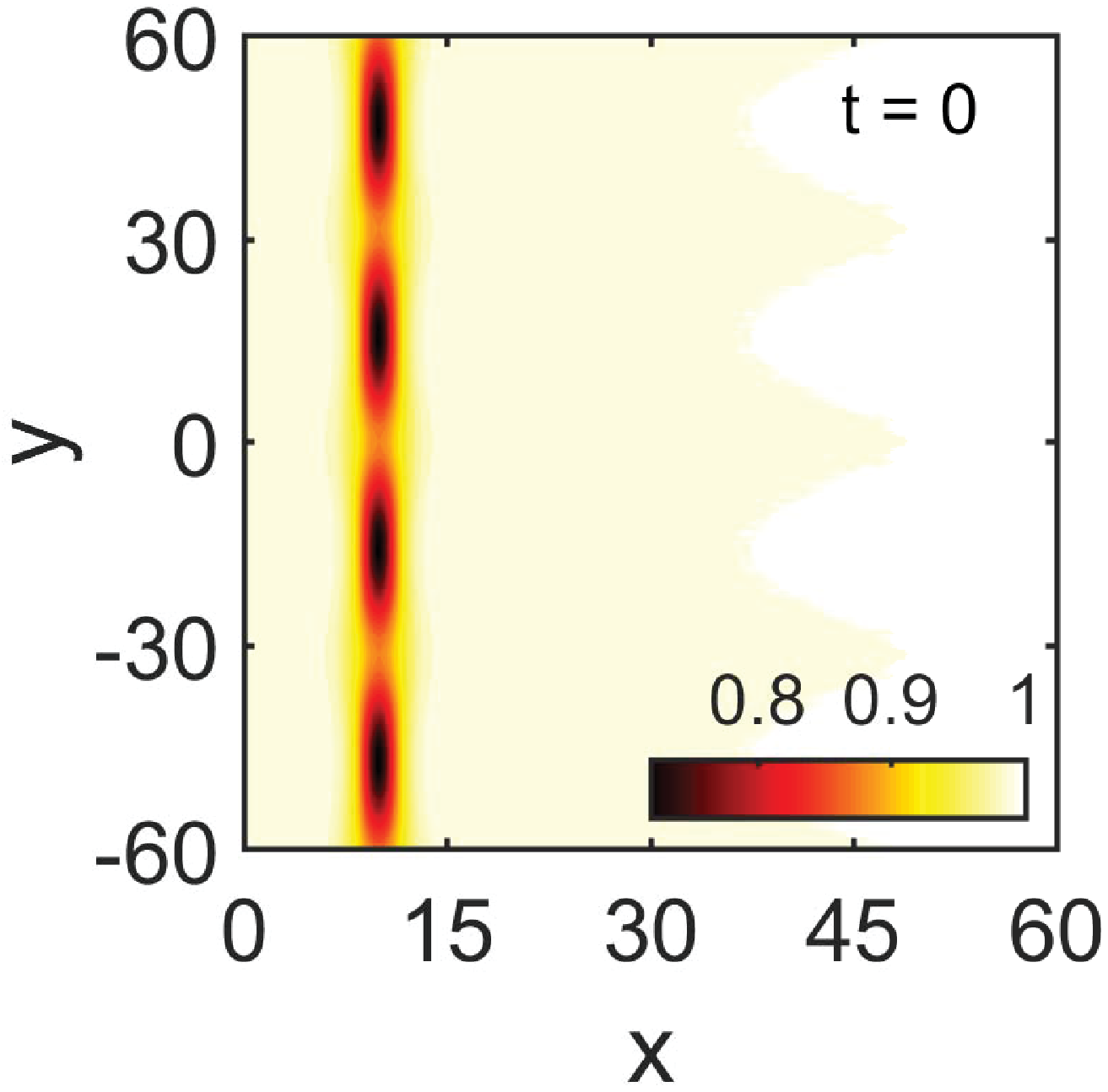}
	\includegraphics[scale=0.3]{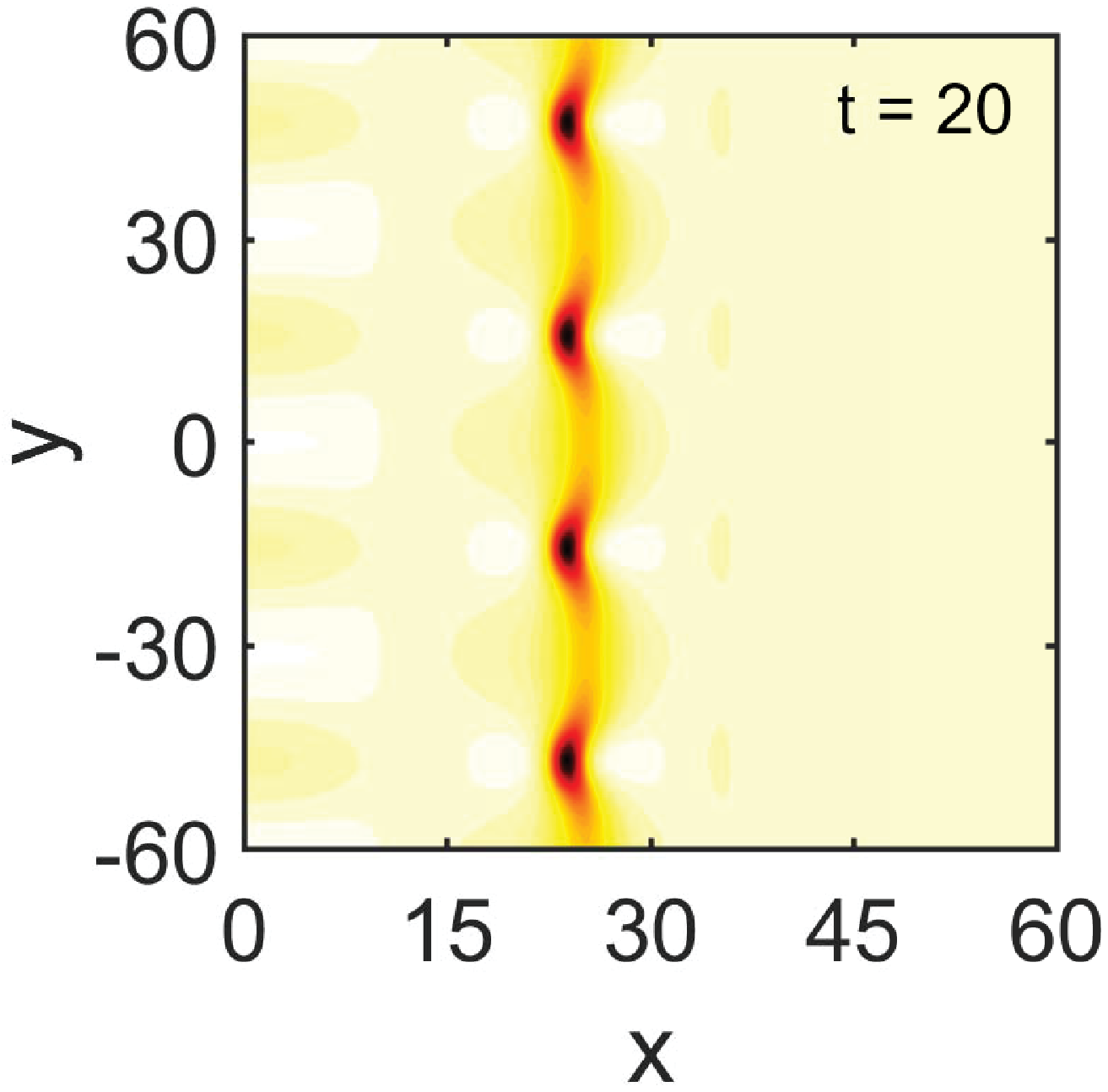}
	\includegraphics[scale=0.3]{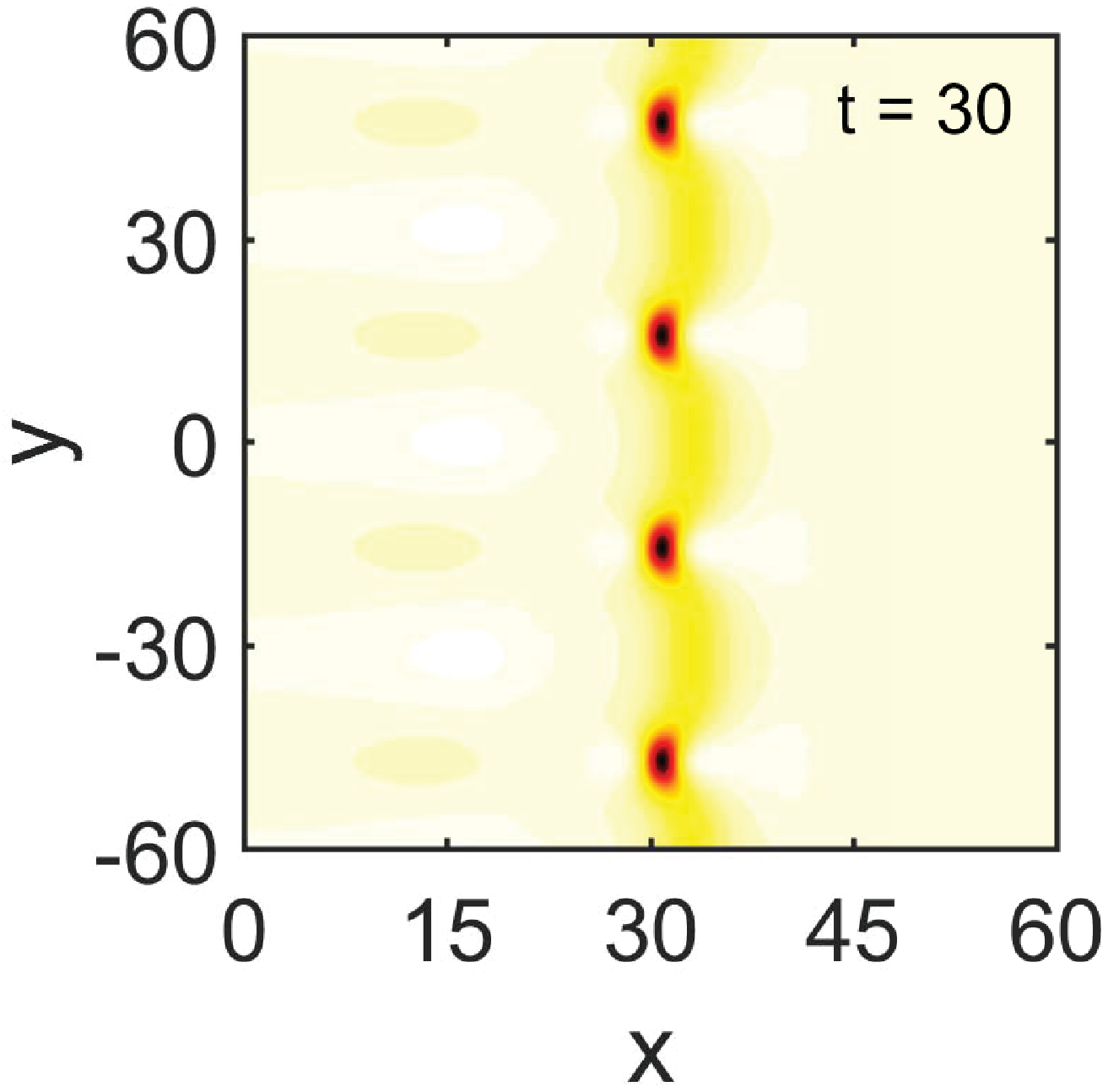}
	\includegraphics[scale=0.3]{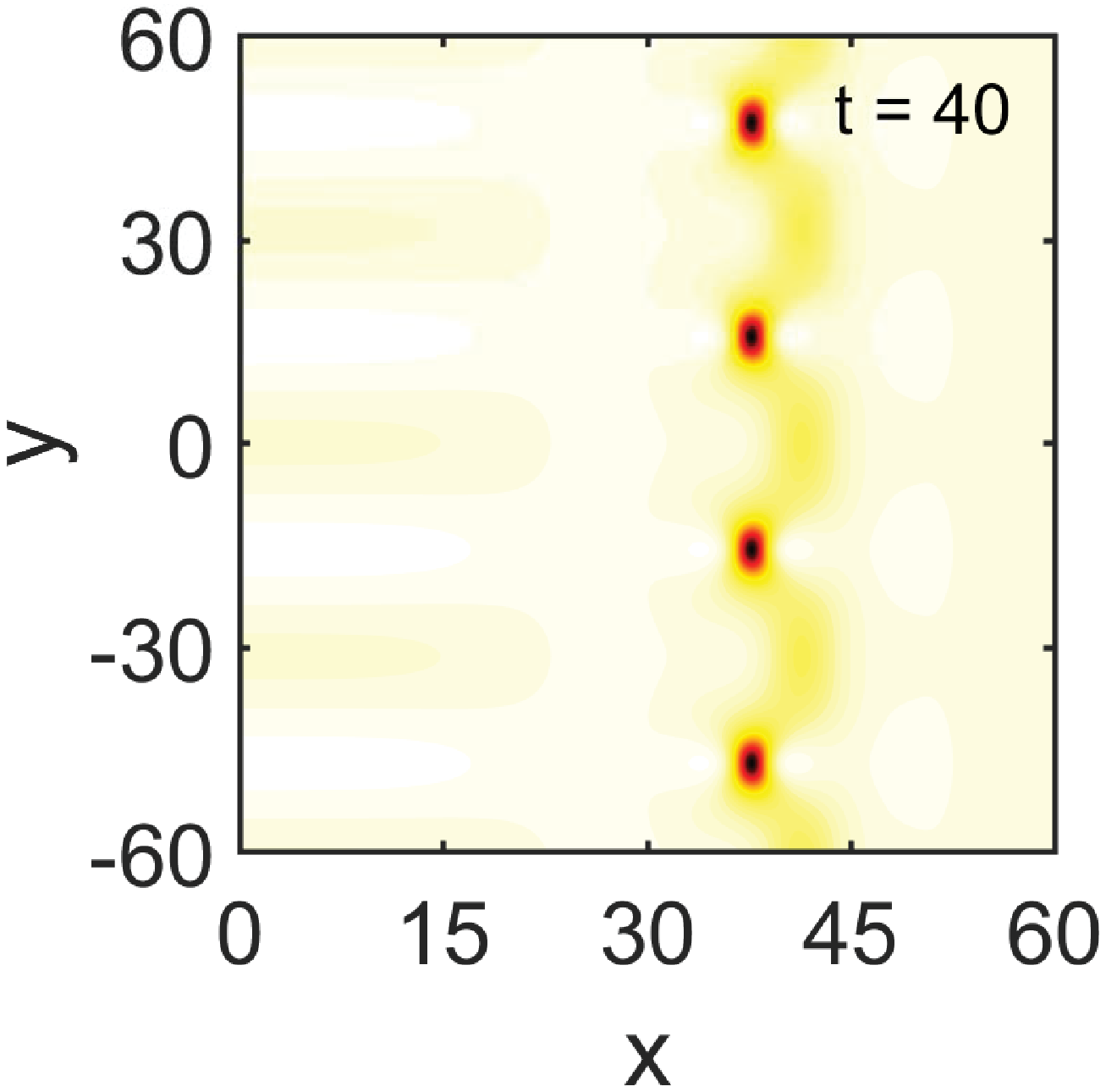}
	\includegraphics[scale=0.3]{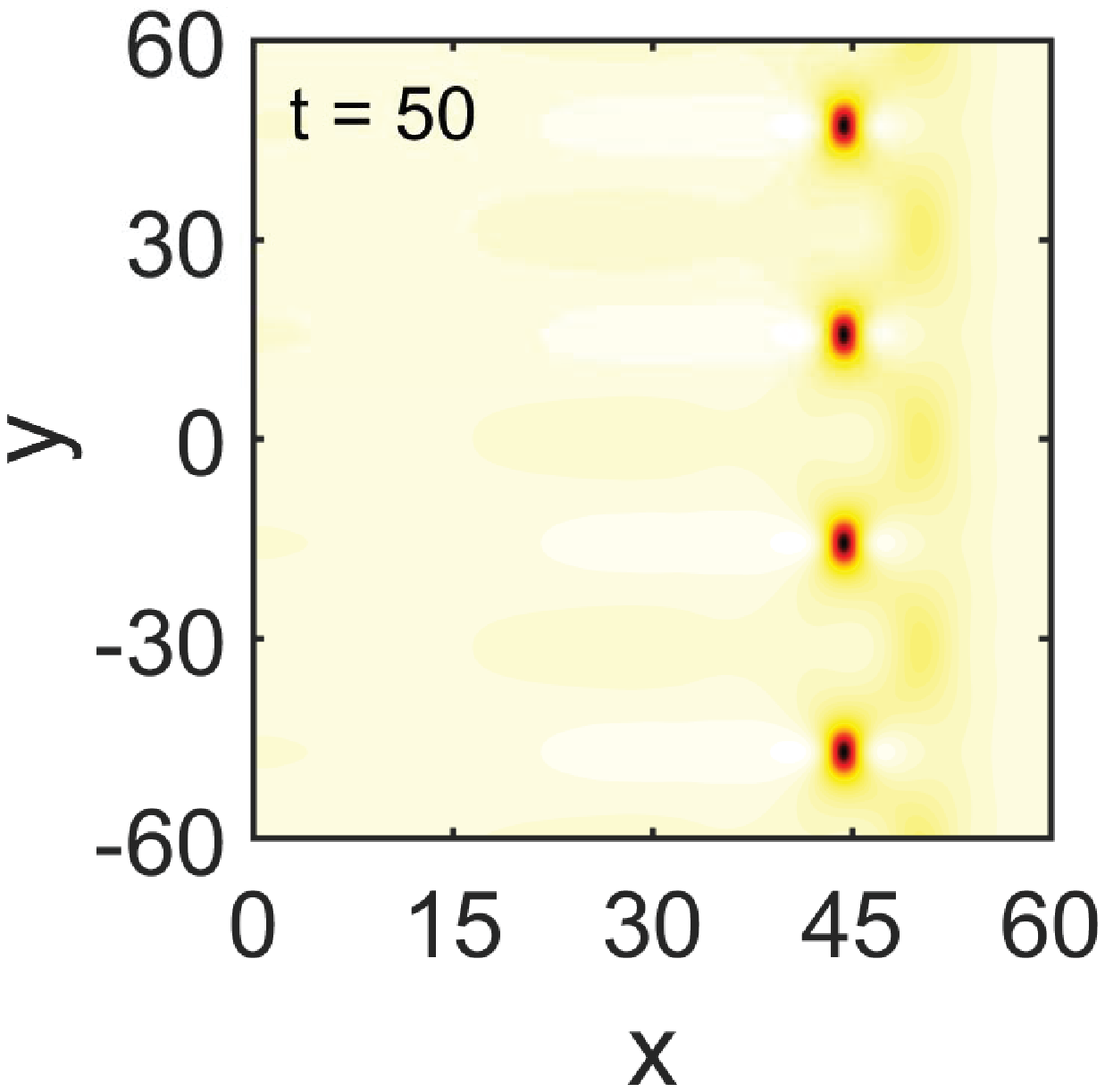}
	\includegraphics[scale=0.3]{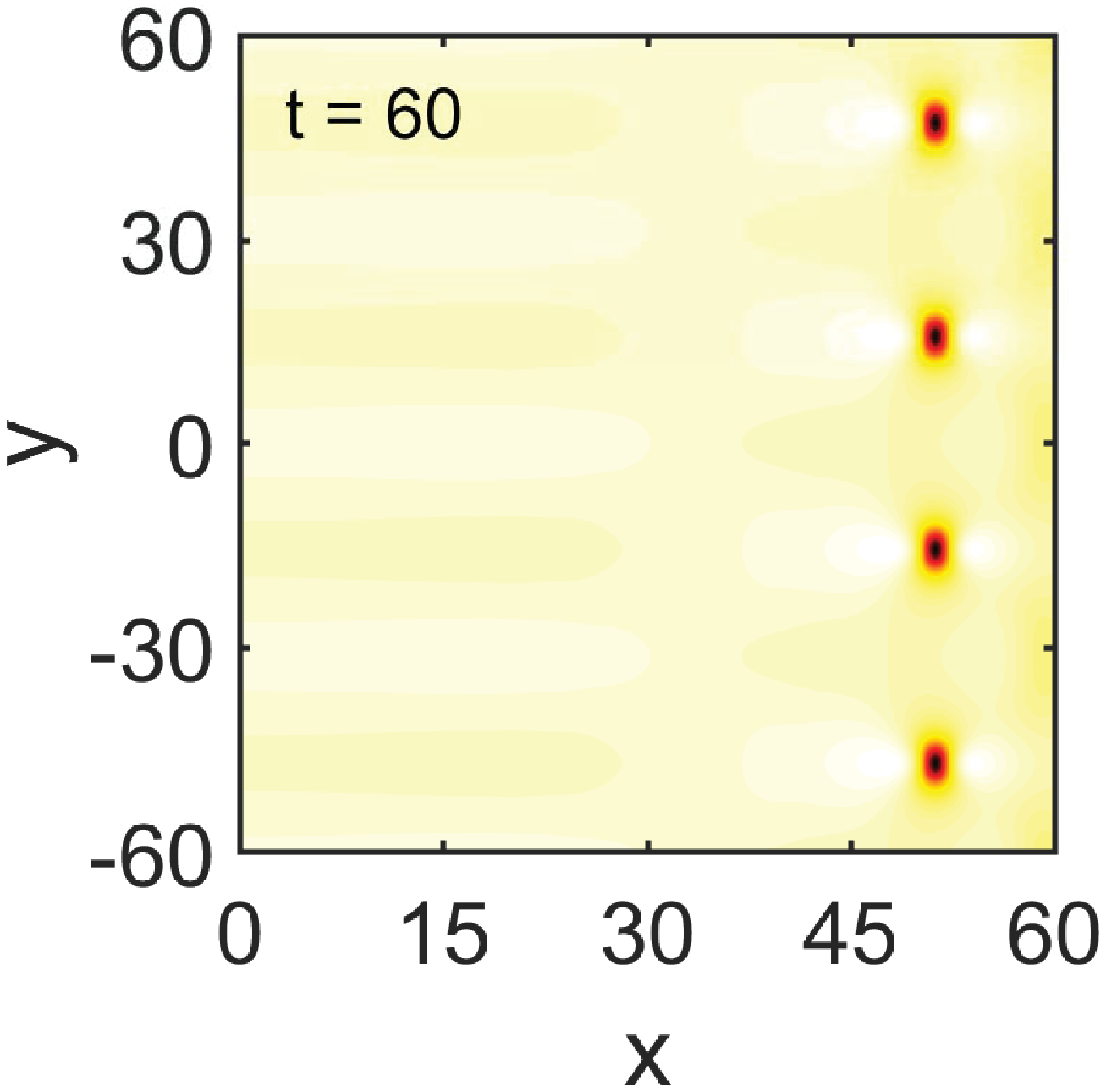}
		\caption{(Color online)	
Same in in Fig.~\ref{fig3} but for a relatively shallow dark soliton stripe.
In this case, after strong undulation, the stripe decays into dark lumps, 
which are formed at approximately $t=50$. 
A zoom of the lowest lump is shown in Fig.~\ref{fig4b}.
All parameters are the same as in Fig.~\ref{fig3} and here
$\kappa=4.5+0.4\cos(0.2y)$.
	}
	\label{fig4}
\end{figure}
%%%%%%%%%%%%%%%%%%%%%%%%%%%%%%%%%%%%%%%%%%%%%%%%%%%%%%%%%%%%%%%%%%%%%%

\begin{figure}[tbp]
	\centering
	\includegraphics[scale=0.25]{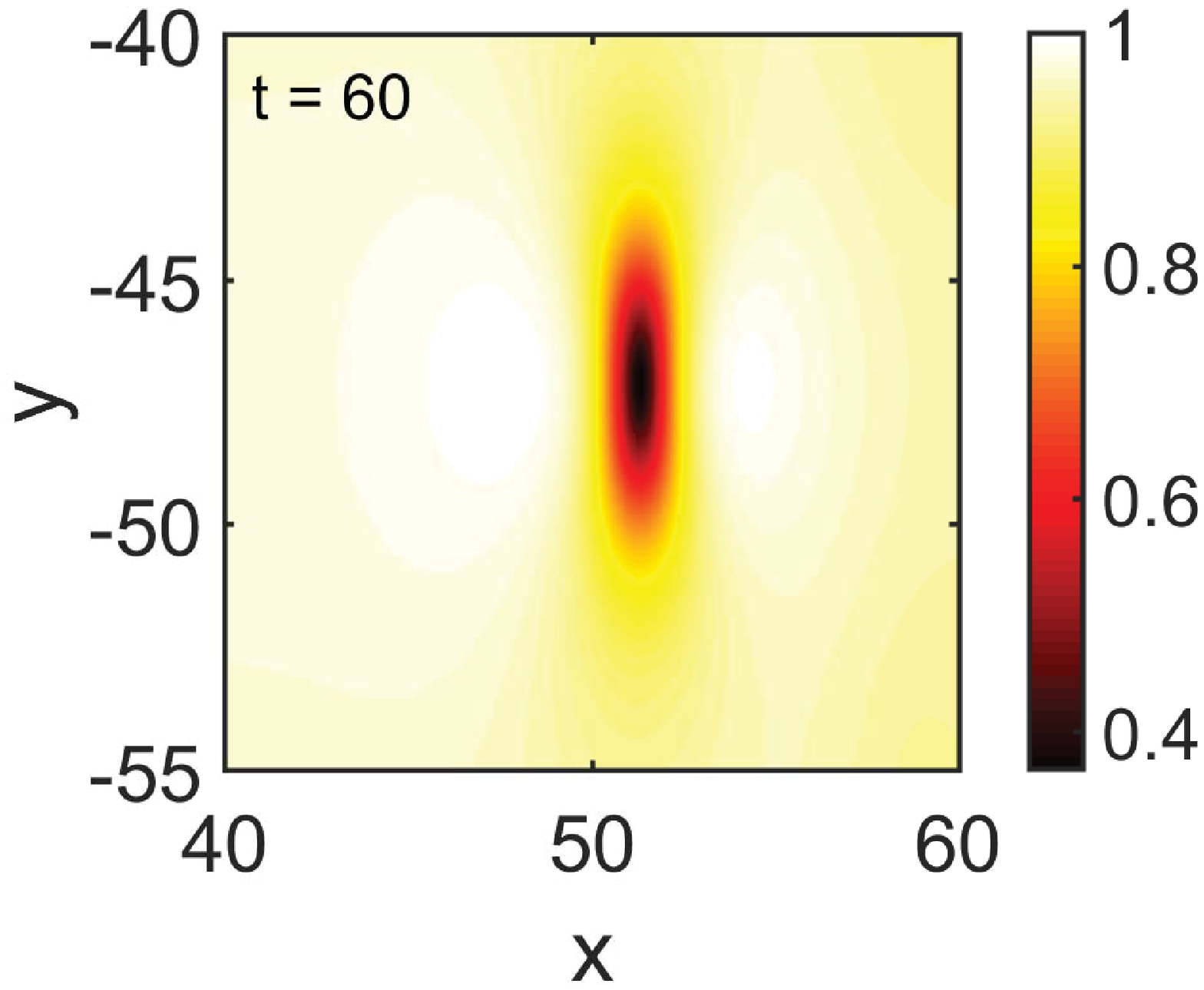}
	\includegraphics[scale=0.25]{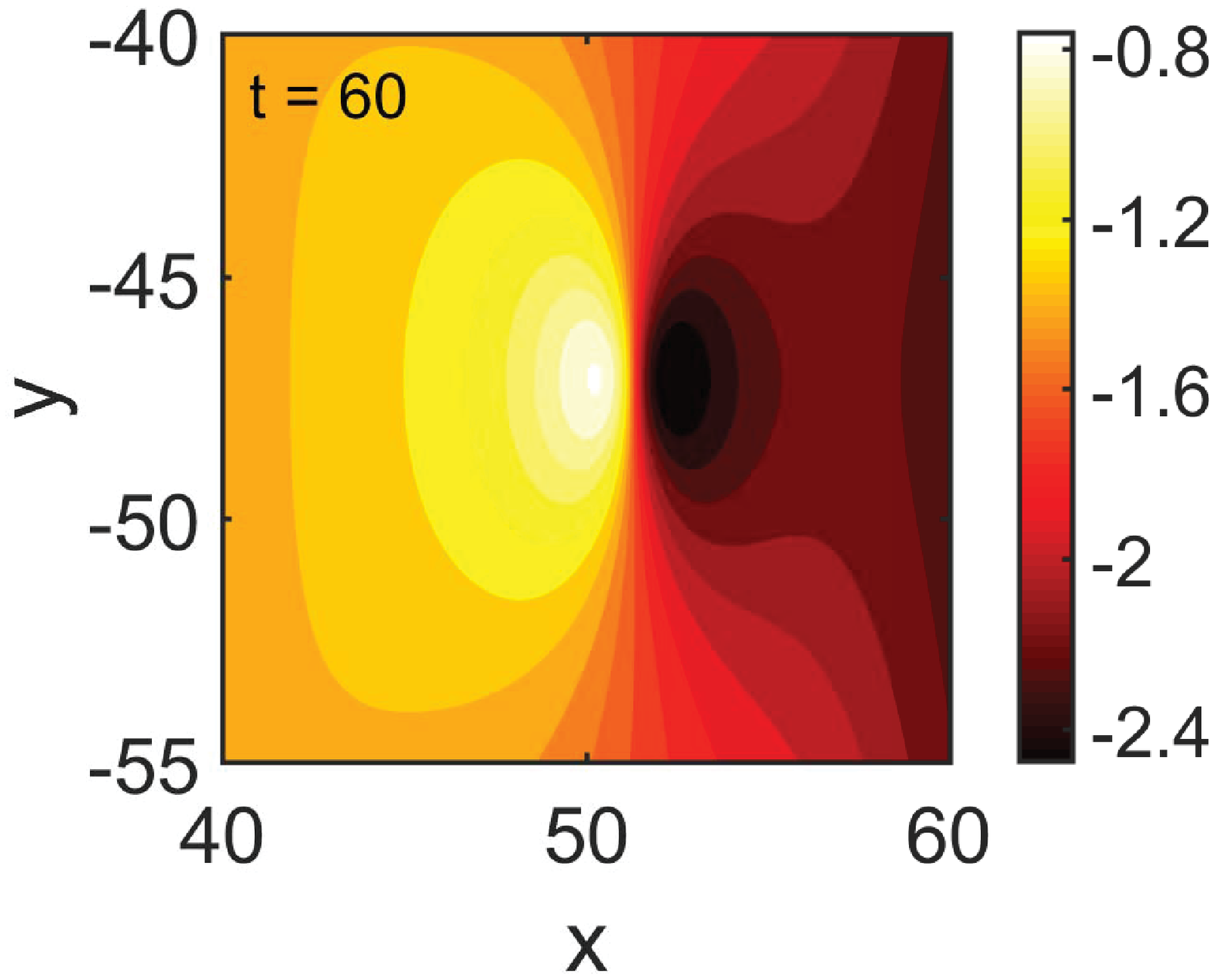}
	\includegraphics[width=0.49\linewidth]{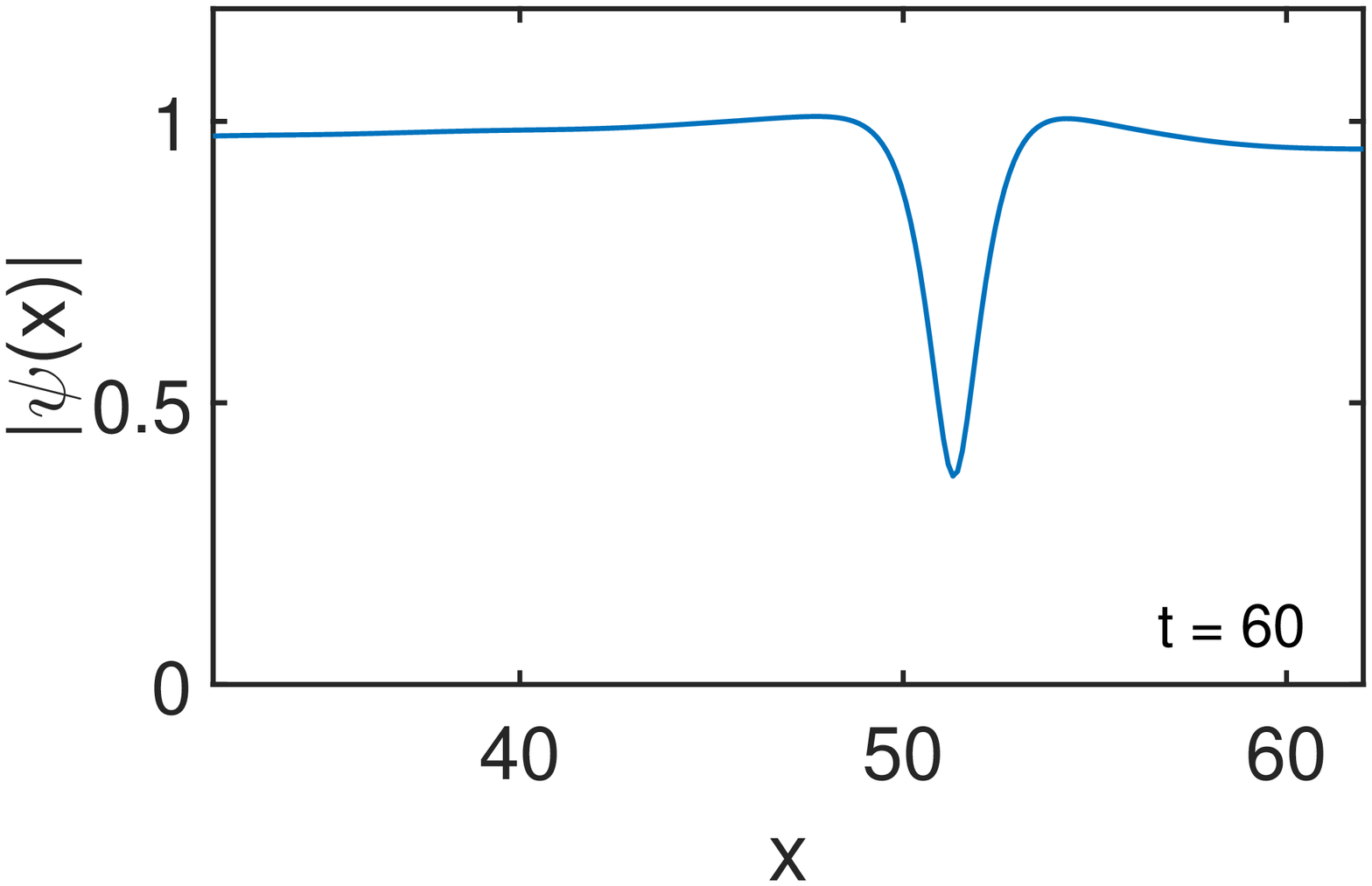}
	\includegraphics[width=0.49\linewidth]{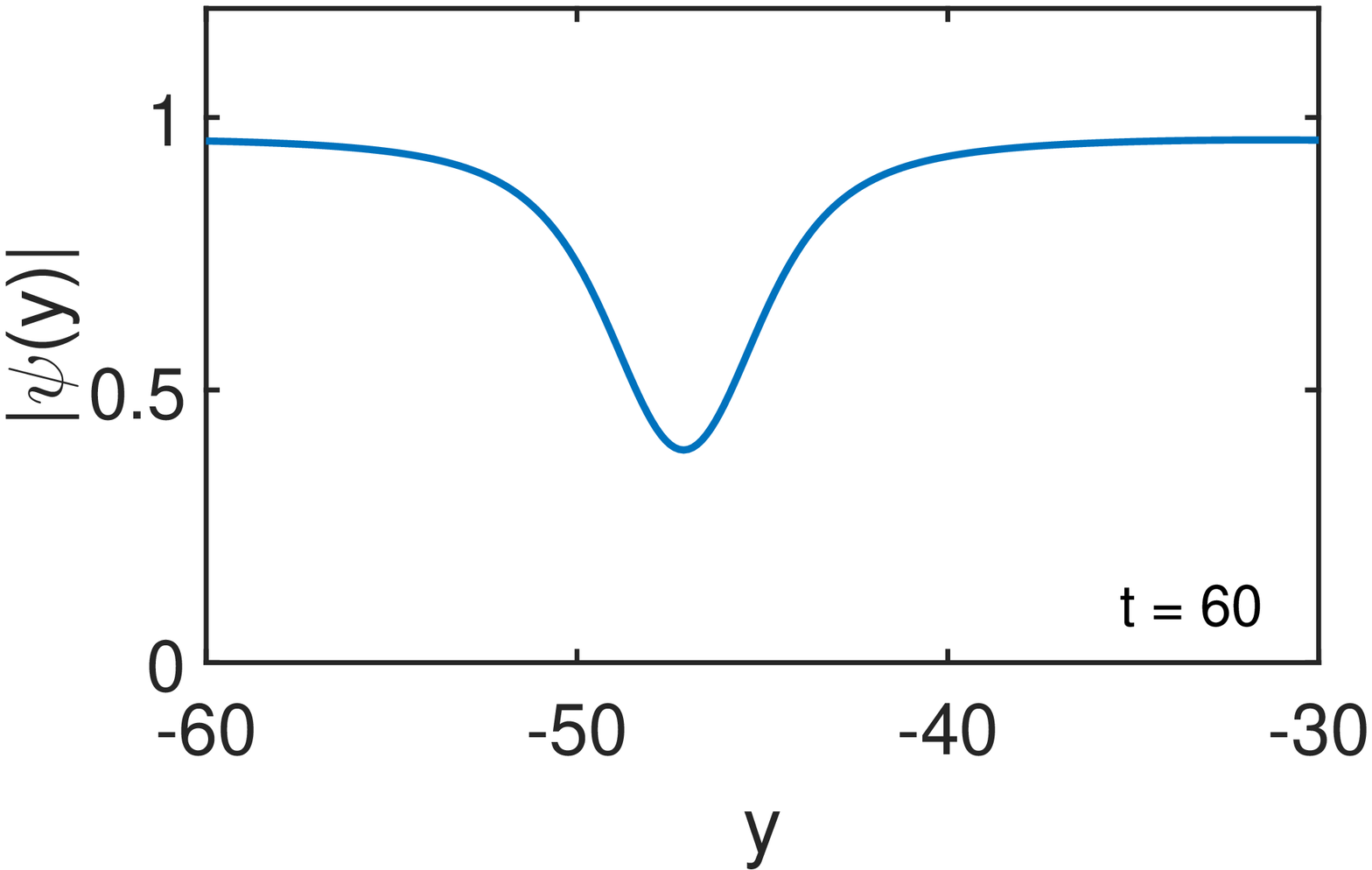}
		\caption{(Color online)	
A zoom depicting the lowest lump 
shown in Fig.~\ref{fig4} at $t=60$. Same layout as in Fig.~\ref{fig3b}.
Notice that the panels clearly reveal the profile of a lump
as depicted in Fig.~\ref{fig1b}.
	}
	\label{fig4b}
\end{figure}

These two scenarios are confirmed by our simulations. Pertinent results are shown in 
Figs.~\ref{fig3} and Fig.~\ref{fig3b} corresponding to relatively deep dark soliton stripes, 
as well as in Figs.~\ref{fig4} and \ref{fig4b} corresponding to shallower dark soliton stripes.   
In both cases, parameter values were chosen as previously and $\epsilon=0.1$.
To accelerate the onset of the snaking instability, we have transversely perturbed
the characteristic parameter of the soliton, $\kappa$, thus using
\begin{equation}
\kappa = \kappa_0 + \kappa_1 \cos(K y),
\label{kperp} 
\end{equation}
where the transverse perturbation wavenumber is set to $K=0.2$, while
$\kappa_0=3$ and $\kappa_1=0.3$ for Figs.~\ref{fig3} and Fig.~\ref{fig3b}, 
whereas $\kappa_0=4.5$ and $\kappa_1=0.4$ for Figs.~\ref{fig4} and Fig.~\ref{fig4b}.

%%%%%%%%%%%%%%%%%%%%%%%%%%%%%%%%%%%%%%%%%%%%%%%%%%%%%%%%%%%%%%%%%%%%%%
%%%%%%%%%%%%%%%%%%%%%%%%%%%%%%%%%%%%%%%%%%%%%%%%%%%%%%%%%%%%%%%%%%%%%%

First, Fig.~\ref{fig3} depicts the evolution of 
a relatively deep dark stripe soliton. It is clearly seen that, after developing strong
undulations, the dark soliton stripe is destroyed, and a chain of 2D
structures, namely vortex-antivortex pairs, are formed
---cf., e.g., the snapshot at $t=50$.
% PGK: In my mind it is not clear at all that the structures are
% time independent traveling ones, nor that they are lumps.
% To show that they are lumps one would need to see if they can be fit
% and that they don't have phase structure (something that I would
% need to see to believe). In fact, I would claim that the
% ``eyes'' of the $t=50$ structures are very much vortices rather than lumps...
In Fig.~\ref{fig3b} we depict a zoom of the lowest vortex pair at $t=60$.
This figure clearly shows that indeed, both the phase plot, which 
is characteristic of a vortex-antivortex pair (top right panel of Fig.~\ref{fig3b}), 
as well as the profiles of the wavefunction's modulus,  
justify the formation of vortex pairs in the deep dark soliton stripe case.

On the other hand, Fig.~\ref{fig4} depicts the evolution of a relatively 
shallow dark stripe soliton. It is observed that, in this case too, the dark soliton stripe 
is destroyed after the onset of the snaking instability. Nevertheless, in this case the 
2D structures that are formed are dark lump solitons. This becomes evident 
in Fig.~\ref{fig4b} depicting the lowest lump in Fig.~\ref{fig4} for $t=60$.
The figure clearly shows that both the wavefunction's modulus and phase, as well as 
the $x$ and $y$ profiles, take a form of a genuine dark lump soliton ---cf.~Fig.~\ref{fig1b} 
for a comparison.

Qualitatively similar results have been obtained with other parameter values 
(results not shown here), a fact that indicates that the dark lump solitons appear 
generically after the onset of the snaking instability of sufficiently weak dark soliton stripes.

\section{Conclusions}
\label{sec:conclu}

Employing multiscale expansions methods we studied the effective hydrodynamic equations 
resulting from a mean-field model for polariton superfluids. The model consists of an 
open-dissipative Gross-Pitaevskii equation for the polariton condensate coupled to a 
rate equation corresponding to the exciton reservoir. We focus on the case of weak 
uniform pumping and sufficiently small polariton loss and stimulated scattering rates.
In particular, we have derived several model equations that are
commonly used in shallow water waves with viscosity ---as well as other physical contexts.
We have thus first derived, at an intermediate stage, a Boussinesq/Benney-Luke type equation,
and then its far-field, a Kadomtsev-Petviashvili-I (KP-I) equation for right- and left-going
waves. By means of the KP-I model, we predict the existence of weakly-localized 
(algebraically decaying) 2D dark-lump solitons. it is found that, in the presence of
dissipation, these dark lumps exhibit a lifetime three times larger than that of
dark soliton stripes. We argued that on the basis of their
robustness (e.g., against transverse undulations) and as a result
of their larger lifetime, these lump structures are likely to be
observable in 2D exciton-polariton superfluids experiments.

Our analytical predictions were corroborated by direct numerical simulations. We found that,
indeed, dark lump solitons do exist and, for sufficiently small values of the formal
perturbation parameter, their dissipative dynamics is well described by the analytical
estimates. Furthermore, we have shown that dark lump solitons, as well as
vortical structures, may emerge spontaneously
after the onset of the snaking instability of sufficiently deep dark soliton stripes.
%
%The fact that dark lumps feature relatively long lifetimes (compared to their planar
%counterparts), and the fact that dark solitons have been observed in a series of experiments
%in polariton superfluids, indicate that dark lump solitons may also be observed in experiments.

It would be interesting to extend our considerations to multi-component (spinor)
polariton superfluid settings ---see, e.g. Refs.~\cite{dsthv2,spinor,berl3,dsth3,li}.
In such settings, a quite relevant investigation would concern the existence of
spinorial, vorticity-free dark lump solitonic structures. It should also be interesting
to use the methodology devised in this work to study other models that are used
in the context of open dissipative systems, such as the Lugiato-Lefever equation \cite{ll}
describing dissipative dynamics in optical resonators. Note also that in
the present setting we have considered homogeneous condensates. However,
parabolic, as well as periodic potentials are routinely used nowadays in
exciton-polariton superfluids. Exploring how such external traps may affect
the present phenomenology would also be an interesting direction for future
work. Such studies are currently in progress and will be reported in future
publications.

\medskip

\acknowledgments
A.S.R.~acknowledges the financial support from FCT through grant SFRH/BSAB/135213/2017.
P.G.K.~gratefully acknowledges the support from NSF-PHY-1602994.
R.C.G.~gratefully acknowledges the support from NSF-PHY-1603058.
J.C.-M.~thanks the financial support from MAT2016-79866-R project (AEI/FEDER, UE).
%

%%%%%%%%%%%%%%%%%%%%%%%%%%%%%%%%%%%%%%%%%%%%%%%%%%%%%

\end{document}